\documentclass[lettersize,journal]{IEEEtran}
\IEEEoverridecommandlockouts
\usepackage{cite}
\usepackage{amsthm}
\usepackage{makecell}
\usepackage{multirow}   
\usepackage{amsmath,amssymb,amsfonts}
\usepackage{graphicx}
\usepackage{textcomp}
\usepackage{xcolor}
\usepackage{amsthm,amsmath,amssymb}
\usepackage{mathrsfs}
\usepackage{amsmath,bm}
\usepackage{algorithm} 
\usepackage{mathtools}
\usepackage{epsf}
\usepackage{subfigure}
\usepackage{cleveref}
\usepackage{verbatim}
\usepackage{algorithmic}
\allowdisplaybreaks[4]
\usepackage{hhline}
\newtheorem{lemma}{Lemma}
\newtheorem{rema}{Remark}
\newtheorem{thm}{Theorem}

\usepackage{caption}

\def\BibTeX{{\rm B\kern-.05em{\sc i\kern-.025em b}\kern-.08em
    T\kern-.1667em\lower.7ex\hbox{E}\kern-.125emX}}
\begin{document}

\title{ {Assistance-Transmission Tradeoff for\\ RIS-assisted Symbiotic Radios}
}

\author{
\IEEEauthorblockN{Hu~Zhou,~Qianqian~Zhang,~Ying-Chang Liang,~\IEEEmembership{Fellow,~IEEE},~and~Yiyang Pei}
\\
\thanks{This work has been submitted to the IEEE for possible publication. Copyright may be transferred without notice, after which this version may no longer be accessible.}
\thanks{
Part of this work was presented in IEEE ICC 2023~\cite{zhou2023}. H. Zhou, Q. Zhang, and Y.-C. Liang are with University of Electronic Science and Technology of China (UESTC), Chengdu 611731, China (e-mails: {huzhou@std.uestc.edu.cn;~qqzhang\_kite@163.com;~liangyc@ieee.org}). Y. Pei is with the Singapore Institute of Technology, 138683, Singapore (e-mail: yiyang.pei@singaporetech.edu.sg).}
}
\maketitle
\vspace{-1cm}
\begin{abstract}
This paper studies the reconfigurable intelligent surface (RIS)-assisted symbiotic radio (SR) system, where an RIS acts as a secondary transmitter to transmit its information by leveraging the primary signal as its RF carrier and simultaneously assists the primary transmission. Conventionally, all reflecting elements of the RIS are used to transmit the secondary signal, which, however, would limit its capability for assisting the primary transmission. To address this issue, we propose a novel RIS partitioning scheme, where the RIS is partitioned into two sub-surfaces, one to assist the primary transmission and the other to transmit the secondary signal. Naturally, there exists a fundamental tradeoff between the assistance and transmission capabilities of RIS regarding the surface partitioning strategy. Considering the coupling effect between the primary and secondary transmissions, we focus on the detection of the composite signal formed by the primary and secondary ones, based on which we propose a novel two-step detector. Then, we formulate the assistance-transmission tradeoff problem to minimize the bit error rate (BER) of the composite signal by jointly optimizing the surface partitioning strategy and the phase shifts of the two sub-surfaces, such that the overall BER of RIS-assisted SR is minimized. By solving this problem, we show that the optimized surface partitioning strategy depends on the channel strength ratio of the direct link to the reflected link. Moreover, performance analysis shows that when the direct link is blocked, exchanging the number of reflecting elements used for assistance and transmission can still achieve almost the same BER of the composite signal thanks to the coupling effect. Finally, extensive simulations show that our proposed RIS partitioning scheme outperforms the conventional schemes which use all reflecting elements for either assistance or transmission.
\end{abstract}


\section{Introduction}
With the rapid emergence of new wireless services and applications such as extended reality and Internet-of-things, it is forecast that the number of connected devices worldwide will increase to over 125 billion by 2030~\cite{giordani2020toward}. Supporting such ultra-massive connections requires huge spectrum and energy resources, which poses significant challenges to 6G~\cite{you2021towards}.
To overcome these challenges, symbiotic radio (SR) has been proposed as a promising solution owing to its spectrum- and energy-efficient features~\cite{you2021towards,chen2020vision}. In SR, the secondary transmitter transmits its information by backscattering the primary signal without requiring additional spectrum and power-hungry RF-chains~\cite{long2019symbiotic}. Meanwhile, the secondary transmission serves as a multi-path to assist the primary transmission since the backscatter link also carries the information of the primary signal in addition to the direct link.
Then, joint detection is employed at the receiver to recover both the primary and secondary signals, which thus enables a mutualistic spectrum-sharing relationship between the primary and secondary transmissions~\cite{liang2020symbiotic,liang2022symbiotic}.

One interesting property of SR is the coupling effect between the primary and secondary transmissions under the maximum likelihood (ML) detection. Specifically, it was observed     in~\cite{yang2018cooperative} that the accurate detection of the primary signal can enhance the accuracy of detecting the secondary signal. In contrast, if the primary signal is detected wrongly, the detection of the secondary signal will also be wrong with a high probability~\cite{yang2018cooperative,liang2020symbiotic}. 
Furthermore, the achievable rate region of SR was characterized in~\cite{liu2018backscatter}  and the mutualistic condition under which the primary and secondary transmissions can benefit each other was analyzed in~\cite{zhang2022mutualistic}. Moreover, considering the coupling effect, resource allocation for SR was studied from the perspective of transmit power minimization~\cite{long2019symbiotic,chu2020resource}, energy efficiency maximization~\cite{yang2021energy}, network utility maximization~\cite{chen2020stochastic}, and sum rate maximization~\cite{guo2019resource,wu2021beamforming}.

Besides SR, reconfigurable intelligent surface (RIS) has recently emerged as another spectrum- and energy-efficient solution for 6G~\cite{wu2019towards,basar2019wireless,liang2019large}. Specifically, RIS is a two-dimensional electromagnetic metasurface equipped with massive low-cost reflecting elements, each of which can introduce a phase shift on the incident signals.
On one hand, by slowly tuning the phase shifts with a frequency $\frac{1}{T_{a}}$, where $T_{a}$ denotes the channel coherence time, RIS can be used to assist an existing system by enhancing the desired signal~\cite{wu2019intelligent,huang2019reconfigurable,han2019large} or mitigating the interference~\cite{ye2022reconfigurable,jiang2022interference} via passive beamforming. On the other hand, by fast tuning the phase shifts with a frequency $\frac{1}{T_{b}}$, where $T_{b}$ denotes the RIS symbol period with $T_{b}\ll T_{a}$, RIS can be used for transmission by modulating its information bits over the incident signals\cite{basar2020reconfigurable,ma2020large}.
Considering the assistance and transmission capabilities of RIS, it is natural to integrate RIS into SR, where RIS acts as a secondary transmitter (STx) of SR to transmit the secondary signal, and simultaneously, the RIS assists the primary transmission by passive beamforming.~\cite{lei2021reconfigurable,liang2022backscatter,yan2020passive,guo2020reflecting,wu2021reconfigurable,li2022reconfigurable,ye2021capacity,zhang2021reconfigurable,zhou2022cooperative,hua2021uav,hua2021novel,wang2021intelligent,hu2020reconfigurable}. In such a system, several modulation schemes have been proposed for the RIS to transmit the secondary signal, such as the backscatter modulation~\cite{liang2022backscatter,yan2020passive}, reflecting modulation~\cite{guo2020reflecting}, and spatial/index modulation~\cite{wu2021reconfigurable,li2022reconfigurable}.
With these modulation schemes, joint active and passive beamforming for RIS-assisted SR has been widely studied to achieve various goals, such as capacity maximization~\cite{ye2021capacity}, transmit power minimization~\cite{zhang2021reconfigurable,zhou2022cooperative}, bit error rate (BER) minimization~\cite{hua2021uav,hua2021novel}, secure broadcasting~\cite{wang2021intelligent}, and weighted sum rate maximization~\cite{hu2020reconfigurable}.

The above works on RIS-assisted SR use all reflecting elements of RIS to transmit the secondary signal. However, such a design may limit the assistance capability of the RIS. This consequently results in limited performance enhancement to the primary transmission as compared to utilizing all reflecting elements of RIS for assistance. In addition, since the secondary signal is modulated over the primary signal, the signal coming from the reflected link is the multiplication of the primary and secondary signals~\cite{liang2022backscatter}. In scenarios where the direct link is blocked, this multiplication relationship between the primary and secondary signals would lead to the ambiguity in the joint decoding of these two signals~\cite{zhang2021reconfigurable}.



Motivated by the above reasons, this paper proposes a novel RIS partitioning scheme for SR, which partitions the RIS into two sub-surfaces, one to assist the primary transmission and the other to transmit the secondary signal. With the partitioning scheme, the ambiguity problem can be effectively addressed since the reflected link via the assistance sub-surface can be viewed as an equivalent direct link. More importantly, the proposed partitioning scheme provides us more degrees of freedom to balance the assistance and transmission capabilities, which naturally reveals a fundamental assistance-transmission tradeoff for RIS in terms of the surface partitioning strategy.
For example, suppose all reflecting elements are assigned to the first sub-surface for assistance. In this case, the performance of the primary transmission can be maximized but the secondary transmission cannot be realized. In contrast, if all reflecting elements are assigned to the second sub-surface for transmission, the RIS beamforming gain is maximized for secondary transmission but the assistance to the primary transmission is limited.
Thus, it is important to find the optimal surface partitioning strategy, under which the best assistance-transmission tradeoff can be achieved.

However, characterizing such a tradeoff is challenging because of the complicated coupling effect between the primary and secondary transmissions in SR. For instance, increasing the number of reflecting elements for assistance does not always enhance the primary transmission at the cost of deteriorating the secondary transmission.
Thus, instead of considering the individual performance of primary and secondary transmissions, which are hard to quantify, we focus on the overall performance of these two transmissions. This is done by exploiting the composite signal received by the receiver, in which the primary and secondary signals are superposed in a both additive and multiplicative manner. 
With the composite signal, we introduce a novel two-step detector that decodes the composite signal first, followed by the mapping from the composite signal to the primary and secondary ones.
The use of the composite signal and the two-step detector allow us to quantify the overall BER performance of the primary and secondary transmissions, which facilitates investigating the assistance-transmission tradeoff in an analytical manner.
Then, we formulate an assistance-transmission tradeoff problem to minimize the BER of the composite signal by jointly optimizing the surface partitioning strategy and phase shifts of RIS.
Although the formulated problem is shown to be a mixed-integer non-linear program, which is difficult to obtain the optimal solution in general, we derive a closed-form solution by first optimizing the phase shifts and then the surface partitioning strategy. Interestingly, it is shown that the optimized surface partitioning strategy under line-of-sight (LoS) channels depends on the channel strength ratio of the direct link to the reflected link. Furthermore, we extend our proposed design and analysis to the cases of higher-order modulation constellations of primary and secondary signals.
In a nutshell, the main contributions of this work are summarized as follows.
\begin{itemize}
    \item We propose a novel RIS partitioning scheme for SR, which partitions the RIS into two sub-surfaces, one to assist the primary transmission and the other to transmit the secondary signal. The proposed scheme provides us more degrees of freedom to balance the assistance and transmission capabilities of RIS, and at the same time eliminates the ambiguity for joint decoding when the direct link is blocked.
    \item We introduce a novel two-step detector which decodes the composite signal first, followed by the mapping from the composite signal to the primary and secondary ones. 
    Then, we formulate the assistance-transmission tradeoff problem to minimize the BER of the composite signal. By solving the problem, we show that the optimized surface partitioning strategy depends on the channel strength ratio of the direct link to the reflected link.
    \item Moreover, we conduct performance analysis to draw uesful insights, including the coupling effect analysis, the performance gain over the conventional RIS design scheme, and the commutative property of assistance-transmission dual capabilities.
    \item Finally, we show by simulations that our proposed RIS partitioning scheme outperforms the conventional schemes which design RIS for either assistance or transmission.
\end{itemize}
\begin{figure*}[t]  
	\centering  
	\setlength{\abovecaptionskip}{-0.05cm}
	\captionsetup{font={scriptsize}}
	\includegraphics[width=5in]{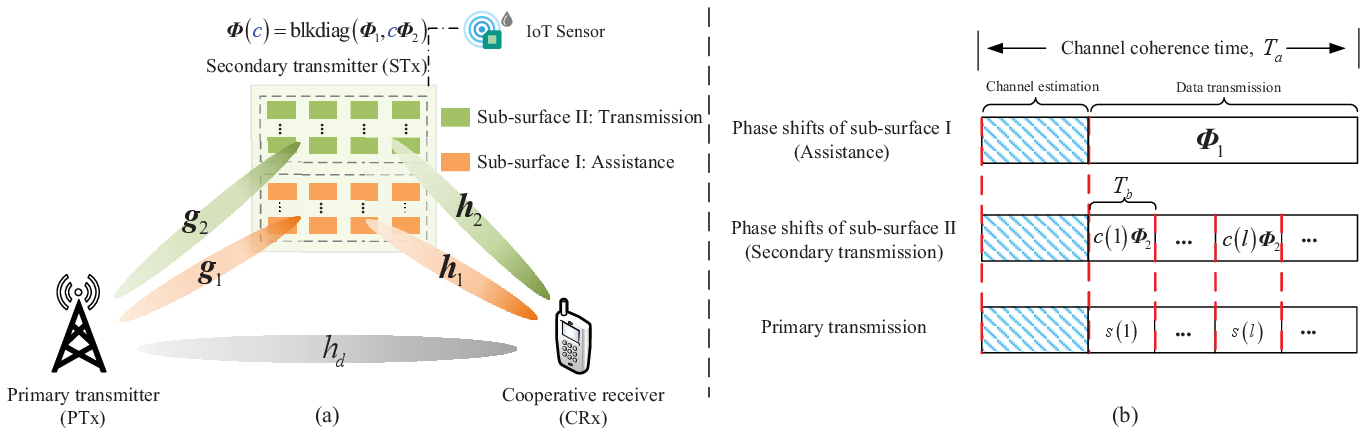}  
	\caption{(a) System model of RIS-assisted SR; (b) The transmission frame structure for RIS-assisted SR.} 
	\label{fig:system-model}  
\end{figure*}

The rest of this paper is organized as follows. Section \ref{sec-system-model} introduces the system model and the proposed RIS partitioning scheme. Section \ref{sec-design- criterion} formulates the assistance-transmission tradeoff problem. Then, Section \ref{sec-optimization-framwork} proposes optimization algorithms to solve the problem. Section \ref{sec-performance-analysis} conducts performance analysis. Section \ref{sec-commutative-property} reveals the commutative property of assistance-transmission dual capabilities. 
Section \ref{sec:-extension} extends the proposed design to the cases of higher-order modulation constellations.
Section \ref{sec-simulation-results} provides simulations to evaluate the proposed scheme. Finally, Section \ref{sec-conslusion} concludes this paper.

Notations: The scalar, vector, and matrix are denoted by the lowercase, boldface lowercase, and boldface uppercase letters, respectively.  $\mathbb{C}^{x \times y} $ denotes the space of $ x \times y $ complex-valued matrices.
$\mathcal{CN}(\mu,\sigma^{2})$ denotes complex Gaussian distribution with mean $\mu$ and variance $\sigma^{2}$. 
$\jmath$ denotes the imaginary unit. $\angle x$, $x^{H}$, and $\Re\{x\}$ denote the phase, conjugate transpose, and real part of number $x$, respectively. $\mathcal{Q}(t)\triangleq\int_{t}^{\infty}\frac{1}{\sqrt{2\pi}}e^{-\frac{1}{2}\eta^{2}}d\eta$ is the complementary distribution function of the standard Gaussian. $\mathbb{N}$ denotes the set of non-negative integers. $\mathrm{blkdiag}(\bm{A},\bm{B})$ is a block-diagonal matrix 
where $\bm{A}$ and $\bm{B}$ located on its main diagonal blocks. $\mathrm{diag}(\bm{t})$ denotes a diagonal matrix whose diagonal elements are the vector $\bm{t}$. $\lceil \cdot \rceil$ denotes the integer ceiling operation.
\section{System Model} \label{sec-system-model}
As illustrated in Fig. \ref{fig:system-model} (a), the RIS-assisted SR system is considered. 
The primary transmitter (PTx) communicates with the cooperative receiver (CRx) with the help from RIS which is equipped with $N$ reflecting elements. Besides, RIS acts as an STx to transmit the secondary signal to the CRx\footnote{Data transmission through an RIS has been experimentally validated.  In \cite{tang2020mimo}, the prototype of RIS-based MIMO-QAM transmitter was implemented, where the transmission rate could be as large as $20 (\mathrm{Mbps})$. As for RIS-assisted SR, an overview of application scenarios is summarized in~\cite{lei2021reconfigurable}, e.g., smart cities and smart homes.}, where the STx refers to the secondary transmitter.
The task of the CRx is to recover both the information from the PTx and the STx via joint decoding. Without loss of generality, we assume that the primary and secondary signals have the same symbol period of $T_{b}$. In the $l$-th symbol period, the primary and secondary signals are denoted by $s(l)\in \mathcal{A}_{s}$ and $c(l)\in \mathcal{A}_{c}$, where $\mathcal{A}_{s}$ and $\mathcal{A}_{c}$ denote the normalized constellation sets of the primary and secondary signals, respectively.
\vspace{-0.4cm}
\subsection{Proposed RIS Design Scheme and Transmission Frame Structure}
In the proposed scheme, RIS is partitioned into two sub-surfaces equipped with $N_{1}$ and $N_{2}$ reflecting elements, respectively, subject to $N_{1}+N_{2}=N$. The sub-surface \uppercase\expandafter{\romannumeral1} is used to purely assist the primary transmission, while the sub-surface \uppercase\expandafter{\romannumeral2} is used to deliver the information of the STx. Herein, the mapping rule between the secondary signal $c(l)$ and the diagonal phase shift matrix of RIS $\bm{\Phi}(c(l)) \in \mathbb{C}^{N \times N}$, is modeled as
\begin{equation} \label{eq: phase-shifts-SAIT}
    \bm{\Phi}(c(l))=\mathrm{blkdiag}(\bm{\Phi}_{1},c(l)\bm{\Phi}_{2}).
\end{equation}

In \eqref{eq: phase-shifts-SAIT}, $\bm{\Phi}(c(l))$ is a block-diagonal matrix with $\bm{\Phi}_{1}$ and $c(l)\bm{\Phi}_{2}$ located on its diagonal blocks; $\bm{\Phi}_{1}=\mathrm{diag}(\phi_{1,1},\cdots,\phi_{1,N_{1}})\in \mathbb{C}^{N_{1} \times N_{1}}$ and $c(l)\bm{\Phi}_{2}\!=\!c(l)\mathrm{diag}(\phi_{2,1},\cdots,\phi_{2,N_{2}})\in\mathbb{C}^{N_{2} \times N_{2}}$
represent the phase shift matrices of sub-surface \uppercase\expandafter{\romannumeral1} and sub-surface \uppercase\expandafter{\romannumeral2}, respectively, where $\bm{\Phi}_{2}$ can be interpreted as the beamforming matrix of the secondary signal $c(l)$. 

To facilitate comprehending the whole system, we briefly introduce the transmission frame structure, as illustrated in Fig. \ref{fig:system-model} (b). Block-flat fading channels are assumed in this paper. For one block of interest, the channel coherence time of $T_{a}$ is divided into two stages. In the first stage, both the PTx and the RIS send pilots, and then the channel state information (CSI) can be estimated via the classical least square method~\cite{swindlehurst2022channel}. In the second stage, the PTx and the RIS start to send the primary and secondary signals to the CRx, respectively.

Note that under our proposed RIS partitioning design, the phase shift matrix of sub-surface \uppercase\expandafter{\romannumeral1}, $\bm{\Phi}_{1}$, remains unchanged during the channel coherence time, while the phase shifts matrix of sub-surface \uppercase\expandafter{\romannumeral2}, $c(l)\bm{\Phi}_{2}$, changes more frequently\footnote{It is reported in~\cite{huang2019reconfigurable} that the energy consumption of phase shifts configuration is small, for example, a $6$-bit resolution phase-shifter consumes $7.8 (\mathrm{mW})$.} with the transmitted symbol $c(l)$ whose symbol period $T_{b}$ is much smaller than the channel coherence time $T_{a}$.
Since this paper focuses on the RIS design in the data transmission stage, we assume perfect CSI is available hereafter. The impact of channel estimation errors during the first stage will be evaluated in Sec. \ref{sec-frame}.


\vspace{-0.4cm}
\subsection{Signal Model}
The channel responses from the PTx to the CRx, from the PTx to the RIS, and from the RIS to the CRx are denoted by $h_{d}\in \mathbb{C}$, $\bm{g}\in\mathbb{C}^{N\times 1}\triangleq[\bm{g}_{1};\bm{g}_{2}]$, and $\bm{h}\in\mathbb{C}^{N\times 1}\triangleq[\bm{h}_{1};\bm{h}_{2}]$, respectively, where $\bm{g}_{1}\in \mathbb{C}^{N_{1}\times 1}$ and $\bm{h}_{1}\in \mathbb{C}^{N_{1}\times 1}$ denote the channel responses associated with the sub-surface \uppercase\expandafter{\romannumeral1}; $\bm{g}_{2}\in \mathbb{C}^{N_{2}\times 1}$ and $\bm{h}_{2}\in \mathbb{C}^{N_{2}\times 1}$ denote the channel responses with sub-surface \uppercase\expandafter{\romannumeral2}.
Let $p_{t}$ denote the PTx transmit power. Then, the received signal at the CRx is given by
\begin{align} \label{eq: received-signal-SAIT}
    y(l)
    &=\underbrace{\sqrt{p_{t}}h_{d}s(l)}_{\text{Direct link}}+
    \underbrace{\sqrt{p_{t}}{\bm{h}_{1}^{H}}\bm{\Phi}_{1}\bm{g}_{1}s(l)}_{\text {Reflected link via sub-surface \uppercase\expandafter{\romannumeral1}} } \nonumber
    \\
    & \quad +
    \underbrace{\sqrt{p_{t}}{\bm{h}_{2}^{H}}\bm{\Phi}_{2}\bm{g}_{2}s(l)c(l)}_{\text{ Reflected link via sub-surface \uppercase\expandafter{\romannumeral2}}}+z(l).
\end{align}
where $z(l)\sim \mathcal{CN}(0,\sigma^{2})$ denotes the complex Gaussian noise at the CRx. Due to the introduction of the proposed RIS partitioning scheme, the signal model is quite different from the existing works which use all RIS elements for either assistance or transmission. Specifically, the CRx receives not only the primary signal from both the direct link and the reflected link via sub-surface \uppercase\expandafter{\romannumeral1}, but also the multiplication of the primary and secondary signals from the reflected link via sub-surface \uppercase\expandafter{\romannumeral2}.

Note that we assume synchronization between the direct signal and the reflected signals, similar to the prior works on symbiotic radio~\cite{hua2021novel,wang2021intelligent,zhang2020capacity}. The reasons are as follows. First, since the RIS only passively reflects the incident signal and there is no receiving module involved, the signal processing delay caused by RIS reflection can be ignored. Second, it is suggested in~\cite{wu2019intelligent} that the RIS should be deployed closer to the base station or the user to achieve a higher performance. Therefore, the distance difference between the direct link and reflected link is small, and thus the propagation delay from the reflected link is negligible. 
In practice, if the delay is non-negligible, it will bring inter-symbol interference for joint detection. In this case,
the classical time synchronization techniques can be applied to cope with the effect of delay~\cite{morelli2007synchronization}.


\begin{rema}
When the symbol duration of the secondary signal is $L$ times larger than that of the primary signal, the symbol duration of the secondary signal can be decomposed into $L$ periods, each of which can be viewed as the scenario where the two signals have the same symbol duration. In this case, our proposed design can be directly applied.
Moreover, our proposed design can be readily extended to the multi-antenna primary transmission. For simplicity, we assume single-antenna transmission to facilitate our design.
\end{rema}

\vspace{-0.5cm}
\subsection{Two-Step Detector Design} \label{sec: system-model-receiver-design}
In RIS-assisted SR, the multiplication relationship between the primary and secondary signals makes it quite challenging to quantify their BER performance~\cite{zhang2022mutualistic}. In this paper, instead of focusing on the 
individual detection of the primary and secondary signals, we focus on the detection of the composite signal, which is composed of the primary and secondary signals. With the composite signal, we propose a novel two-step detector, which enables us to quantify both the overall and individual BER performance of the primary and secondary signals and to study the assistance-transmission tradeoff analytically.
Specifically, the composite signal is defined as
\begin{align} \label{eq: composite-signal-SAIT}
    x(l)=(h_{d}+{\bm{h}_{1}^{H}}\bm{\Phi}_{1}\bm{g}_{1}+{\bm{h}_{2}^{H}}\bm{\Phi}_{2}\bm{g}_{2}c(l))s(l),
\end{align}
where 
$x(l)$ is a function of $s(l)$ and $c(l)$.
With $x(l)$, the received signal at the CRx can be rewritten as 
\begin{align} \label{eq:received-signal-JAIT-re}
    y(l)=\sqrt{p_{t}}x(l)+z(l).
\end{align}

From \eqref{eq:received-signal-JAIT-re}, the composite signal affects the overall BER performance of RIS-assisted SR and provides us with a new view on the detector design. Specifically, we can first decode the composite signal from the received signal. Then, according to the mapping rule between $x(l)$ and $\{s(l),c(l)\}$, we recover $\{\hat{s}(l),\hat{c}(l)\}$ from the decoded composite signal $\hat{x}(l)$. Overall, the above two-step detection is summarized as follows
\begin{align} 
&\text{\emph{Step 1}}: \hat{x}(l)=\arg \min_{x(l)\in\mathcal{A}_{x}} \left\| y(l)-\sqrt{p_{t}}x(l)\right\|^{2},\label{eq-ML-Detector-x-SAIT}\\
&\text{\emph{Step 2}}:\left\{\hat{s}(l),\hat{c}(l)\right\}\xleftarrow{\text{bit mapping}}\hat{x}(l).\label{eq-x-sc-SAIT}
\end{align}
In \eqref{eq-ML-Detector-x-SAIT}, $\mathcal{A}_{x}$ is the constellation set of $x(l)$. 
Note that given the constellation sets of $\mathcal{A}_{s}$ and $\mathcal{A}_{c}$, the phase shifts of RIS, and the channel coefficients, $A_x$ is a finite set.

From \eqref{eq-ML-Detector-x-SAIT} and \eqref{eq-x-sc-SAIT}, it is observed that the overall BER of $s(l)$ and $c(l)$ are related to that of $x(l)$. According to \emph{Step 1},
to guarantee a good BER performance of $s(l)$ and $c(l)$,
the CRx needs to first decode $x(l)$ accurately.
Then, in terms of \emph{Step 2}, a good bit mapping rule of the composite signal, e.g., Gray mapping, would further decrease the BER of the primary and secondary signals. Inspired by these two observations, we will design the surface partitioning strategy and the phase shifts of the two sub-surfaces to maximize the overall BER performance in the next section.
\vspace{-0.4cm}
\section{RIS Design for Symbiotic Radio} \label{sec-design- criterion}
In this section, we present the design criteria to guide RIS design, and then formulate an assistance-transmission tradeoff problem.
As an example, we assume the primary and secondary signals adopt quadrature phase shift keying (QPSK) and binary phase shift keying (BPSK), respectively. The extension to the higher-order modulation constellations will be studied in Sec. \ref{sec:-extension}. 
In addition, for notational simplicity, the symbol period notation $l$ is dropped hereafter.
\vspace{-0.4cm}
\subsection{Design Criteria}
\label{sec-Design-criteria}
According to the proposed two-step detector, it is found that the minimum Euclidean distance and the Hamming distance between neighboring constellations of the composite signal have a dominant impact on the performance of \emph{Step 1} and \emph{Step 2} of the proposed detector, respectively~\cite{proakis2001digital,tse2005fundamentals}, which are elaborated as follows.
\subsubsection{Minimum Euclidean Distance (Step 1 of Two-Step Detector)}
The larger minimum Euclidean distance helps distinguish nearest constellations more accurately, which thus leads to better symbol error rate (SER) performance of the composite signal in the \emph{Step 1} of the proposed detector. Hence, we aim to maximize the minimum Euclidean distance, given by 
\begin{align}
    D_{\min}\triangleq\min\limits_{m\neq k}\{D_{m,k},m,k=1,\cdots,|\mathcal{A}_{x}|\},
\end{align}
where $D_{m,k}$ is the Euclidean distance between any two constellations, defined as $D_{m,k}=|x_{m}-x_{k}|$, $\forall x_{m},x_{k}\in \mathcal{A}_{x}$.
\subsubsection{The Hamming Distance Between Neighboring Constellations (Step 2 of Two-Step Detector)} \label{sec:Hamming-distance}
Different Hamming distances lead to different bit-mapping rules, which affect the overall BER performance of primary and secondary signals in the \emph{Step 2} of the proposed detector.
Considering the classical Gray mapping, we assume that the constellations of $s$ and $c$ are mapped into bit sequences $\{00,01,11,10\}$ and $\{1,0\}$, respectively. Then, the composite signal $x$ is of $3$ bits whose first two bits denote the primary signal $s$ and the last bit denotes the secondary signal $c$. Letting $h_{e}=h_{d}+{\bm{h}_{1}^{H}}\bm{\Phi}_{1}\bm{g}_{1}$, the composite signal can be rewritten as $x=(h_{e}+{\bm{h}_{2}^{H}}\bm{\Phi}_{2}\bm{g}_{2}c)s$ according to \eqref{eq: composite-signal-SAIT}. In this way, the composite signal can be regarded as two rotated versions of $s$, where the phase rotation is denoted by $\angle(h_{e}\!+\!{\bm{h}_{2}^{H}}\bm{\Phi}_{2}\bm{g}_{2}c)$, $\forall c\!\in\! \mathcal{A}_{c}\!=\!\{1,-1\}$.
\begin{figure}[t]  
	\centering  
	\setlength{\abovecaptionskip}{-0.05cm}
	\captionsetup{font={scriptsize}}
	\includegraphics[width=3.2in]{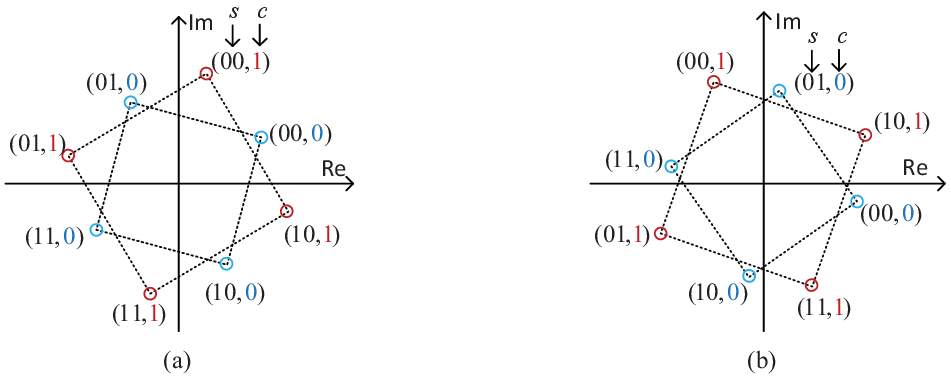}
	\caption{Bit mapping rules: $x\rightarrow (s,c)$ with $s$ being QPSK, $c$ being BPSK, and $x$ being two rotated versions of $s$ when $c$ takes $\mathcal{A}_{c}=\{1,-1\}$. (a) Example of the bit mapping rule when $|\angle(h_{e}\!+\!{\bm{h}_{2}^{H}}\bm{\Phi}_{2}\bm{g}_{2}c)|\leq \frac{\pi}{4}$;
	(b) Example of bit mapping rule when $|\angle(h_{e}\!+\!{\bm{h}_{2}^{H}}\bm{\Phi}_{2}\bm{g}_{2}c)|> \frac{\pi}{4}$.}
	\label{fig:bit-mapping}
\end{figure}

Then, depending on whether the absolute value of the phase rotation, $|\angle(h_{e}\!+\!{\bm{h}_{2}^{H}}\bm{\Phi}_{2}\bm{g}_{2}c)|$, is greater than $\frac{\pi}{4}$, the bit mapping rules of the composite signal can be divided into two categories by assuming $\angle h_{e}=0$. On one hand, if $|\angle(h_{e}\!+\!{\bm{h}_{2}^{H}}\bm{\Phi}_{2}\bm{g}_{2}c)|\leq \frac{\pi}{4}$, then the rotated versions of the primary signal still locate in the decision region of the primary signal, which corresponds to the bit mapping rule shown in Fig. \ref{fig:bit-mapping} (a). Here, we assume that the decision regions of the primary signal with QPSK constellations are the four quadrants, respectively.
For instance, the two rotated constellations with bit sequences $(00,0)$ and $(00,1)$ are still within the first quadrant, which is the decision region of the primary signal with bit sequence $(00)$. In this case, the Hamming distance between the constellations in the same quadrant is $1$ bit, while the Hamming distance between the neighboring constellations in the different quadrants is $2$ bits. On the other hand, if $|\angle(h_{e}\!+\!{\bm{h}_{2}^{H}}\bm{\Phi}_{2}\bm{g}_{2}c)|> \frac{\pi}{4}$, the rotated versions of the primary signal locate outside the decision region of the primary signal, which is illustrated by the example shown in Fig. \ref{fig:bit-mapping} (b). Through comparison, it is easy to find that in case the rotated versions still locate within the decision region of the primary signal, the Hamming distance between neighboring constellations would be smaller. Then, considering the general case where $\angle (h_{e})=\angle(h_{d}+{\bm{h}_{1}^{H}}\bm{\Phi}_{1}\bm{g}_{1})\neq 0$, the phase rotation constraints can be expressed as $\mathrm{C1}$ and $\mathrm{C2}$ when $c$ takes $1$ and $-1$, respectively.
\begin{align}
    &\mathrm{C1}: 0\leq |\angle(h_{e}+\bm{h}_{2}^{H}\bm{\Phi}_{2}\bm{g}_{2})-\angle(h_{e})|\leq \frac{\pi}{4}, 
    \\
   & \mathrm{C2}:0\leq |\angle(h_{e}-\bm{h}_{2}^{H}\bm{\Phi}_{2}\bm{g}_{2})-\angle(h_{e})|\leq \frac{\pi}{4}.
\end{align}

\subsection{Assistance-Transmission Tradeoff} \label{sec-assistance-transmission-tradeoff}
{In the proposed RIS partitioning scheme, it is important to determine how many reflecting elements should be allocated for assistance and transmission. From \eqref{eq: received-signal-SAIT}, if all reflecting elements are allocated for assistance, i.e., $N_1=N$, $N_2=0$, the received signal becomes $y=(h_{d}+{\bm{h}_{1}^{H}}\bm{\Phi}_{1}\bm{g}_{1})s+z$. In this case, $\bm{\Phi}_{1}$ can be designed to maximize the BER performance of primary transmission while the secondary signal cannot be transmitted. On the contrary, if all reflecting elements are allocated for transmission, i.e., $N_1=0$, $N_2=N$, the received signal becomes $y=(h_{d}+c{\bm{h}_{2}^{H}}\bm{\Phi}_{2}\bm{g}_{2})s+z$. In this case, the beamforming gain is maximized for the secondary signal $c$. However, due to the variation of $c$ for secondary data transmission, the composite channel $h_{d}+c{\bm{h}_{2}^{H}}\bm{\Phi}_{2}\bm{g}_{2}$ for the primary transmission is also varied, which leads to limited assistance for primary transmission as compared to the case where the composite channel is $h_{d}+{\bm{h}_{1}^{H}}\bm{\Phi}_{1}\bm{g}_{1}$. Thus, there exists an assistance-transmission tradeoff in terms of the surface partitioning strategy.}
This paper aims to achieve the best tradeoff by minimizing the overall BER of the primary and secondary signals, which can be realized by minimizing the BER of the composite signal. Then, with the above design criteria, our objective is to maximize the minimum Euclidean distance of the composite signal subject to its Hamming distance constraints, unit-modulus phase shift constraints, and the total number constraint of the RIS reflecting elements, by jointly optimizing the surface partitioning strategy and the RIS phase shifts. Mathematically, the assistance-transmission tradeoff problem is formulated as\footnote{
When the primary and secondary receivers are spatially separated, our proposed RIS partitioning design is still applicable, while in this case,  the assistance-transmission tradeoff depends on not only the surface partitioning strategy but also the RIS beamforming gain towards the directions of the primary and secondary receivers.}
\begin{subequations}
	\begin{align}
	\underline{\textbf{\text{P1:}}} \quad 
	&\max \limits_{\bm{\Phi}_{1},\bm{\Phi}_{2},N_{1},N_{2}}\! D_{\min}\!\triangleq\!\min\limits_{m \neq k}\left\{D_{m,k},m,k\!=\!1,\cdots,|\mathcal{A}_{x}|\right\}\\
	&\ \mbox{s.t.} \ \: \mathrm{C1,C2}, \\
	& \quad \quad\mathrm{C3}: N_{1}+N_{2}=N, N_{1},N_{2}\in \mathbb{N},\\
	& \quad \quad\mathrm{C4}: |[\bm{\Phi}_{1}]_{n,n}|=1, n = 1,\cdots,N_{1},\\
	& \quad \quad \mathrm{C5}: |[\bm{\Phi}_{2}]_{n,n}|=1, n = 1,\cdots,N_{2},
	\end{align}
\end{subequations}
where $\mathrm{C3}$ denotes the total number constraint of the RIS reflecting elements; $\mathrm{C4}$ and $\mathrm{C5}$ denote the phase shifts constraints associated with the two sub-surfaces. 

However, \textbf{P1} is difficult to be solved due to the following reasons. First, the phase rotation constraints ($\mathrm{C1}$, $\mathrm{C2}$) and the phase shifts constraints ($\mathrm{C4}$, $\mathrm{C5}$) are non-convex. Second, the constraint $\mathrm{C3}$ restricts $N_{1}$ and $N_{2}$ to be non-negative discrete values, which involves integer programming. Overall, \textbf{P1} is shown to be a mixed-integer non-linear program. Generally speaking, there is no standard method to directly obtain the optimal solution. Nevertheless, in the following, we propose an optimization framework to solve this problem.
\vspace{-0.4cm}
\section{Optimization Framework} \label{sec-optimization-framwork}
In this section, we solve \textbf{P1} by first optimizing the phase shifts of the two sub-surfaces under any surface partitioning strategy. Then, with the obtained phase shifts, we optimize the surface partitioning strategy to directly solve \textbf{P1} without iterations between the phase shifts optimization and surface partitioning strategy optimization\footnote{In fact, the optimal solution to \textbf{P1} should be obtained by solving it for all the variables at once, which, however, is challenging. Therefore, we consider solving \textbf{P1} by dividing it into two subproblems in order to obtain some design guidelines. The obtained solution can serve as a suboptimal solution to \textbf{P1}.}. 
\vspace{-0.3cm}
\subsection{Optimization of RIS Phase Shifts} \label{sec: phase shifts optimization}
Given the RIS partitioning strategy, the problem \textbf{P1} can be recast as the following problem.
\begin{subequations} \nonumber
	\begin{align}
	\underline{\textbf{\text{P2:}}} \quad 
	&\max \limits_{\bm{\Phi}_{1},\bm{\Phi}_{2}}\ D_{\min}\triangleq\min\limits_{m \neq k}\left\{D_{m,k},m,k=1,\cdots,|\mathcal{A}_{x}|\right\}\\
	&\ \mbox{s.t.} \ \: \mathrm{C1,C2,C4,C5}. 
	\end{align}
\end{subequations}

Although simplified, \textbf{P2} is still challenging to be solved due to the complicated expressions of $\mathrm{C1}$-$\mathrm{C2}$. However, it is observed that we can first solve \textbf{P2} without considering $\mathrm{C1}$-$\mathrm{C2}$ to obtain a closed-form solution of the phase shifts. Then, we can adjust the partitioning strategy to satisfy the phase rotation constraints ($\mathrm{C1},\mathrm{C2}$) in the next subsection. 
Now, we derive the closed-form solution to  \textbf{P2} without considering $\mathrm{C1}$ and $\mathrm{C2}$, which is shown in Theorem \ref{thm-phase-shifts}.
\vspace{-0.1cm}
\begin{thm} \label{thm-phase-shifts}
The optimized phase shifts $\bm{\phi}_{1}^{\star}$ and $\bm{\phi}_{2}^{\star}$ of the sub-surface \uppercase\expandafter{\romannumeral1} and sub-surface \uppercase\expandafter{\romannumeral2}, which could maximize the minimum Euclidean distance of the composite signal, are given by
\begin{align}
    \phi_{1,n}^{\star}&=e^{\jmath(\angle(h_{d})-\angle(f_{1,n}))},\forall n= 1,\cdots,N_{1}, \label{eq: phase-shifts-I}
    \\
    \phi_{2,n}^{\star}&=e^{\jmath(\overline{\phi}-\angle(f_{2,n}))},\forall n= 1,\cdots,N_{2}, \label{eq: phase-shifts-II}
\end{align}
where $\overline{\phi}$ denotes the additional phase shift imposed on $\bm{\phi}_{2}$,
\begin{align} \label{eq:-common-phase}
    \overline{\phi}=\arg \max\limits_{\overline{\phi}\in [0,2\pi) } D_{\min} = \angle(h_{d})+\frac{\pi}{4} \ \mathrm{or} \ \angle(h_{d})+\frac{3\pi}{4}.
\end{align}
\end{thm}
\begin{IEEEproof}
Please refer to Appendix \ref{appendix-phase-shifts} for details.
\end{IEEEproof}
\begin{rema}
When the sub-surface \uppercase\expandafter{\romannumeral2} transmits the secondary signal, its phase shifts will be changed, which may lead to a large fluctuation in the energy of the received signal. Therefore, compared with the sub-surface \uppercase\expandafter{\romannumeral1} which is used to assist the primary transmission, the phase shifts of sub-surface \uppercase\expandafter{\romannumeral2} require an additional common phase shift $\overline{\phi}$ to reduce such a fluctuation, where $\overline{\phi}$ is determined by \eqref{eq:-common-phase}.
Notably, adding such a common phase shift to the sub-surface \uppercase\expandafter{\romannumeral2} does not change its passive beamforming gain. 
\end{rema}
\vspace{-0.4cm}
\subsection{Optimization of Surface Partitioning Strategy}
Recall that the optimized phase shifts are obtained by relaxing the constraints $\mathrm{C1}$-$\mathrm{C2}$. With the optimized RIS phase shifts, we can rewrite the phase rotation constraints $\mathrm{C1}$-$\mathrm{C2}$ as $\mathrm{C1'}$ and adjust the surface partitioning strategy to satisfy it, as shown in the following Lemma.
\begin{lemma} \label{lem-constraint}
Based on Theorem \ref{thm-phase-shifts} and constraints $\mathrm{C1}$-$\mathrm{C2}$, the condition on the surface partitioning variables (i.e., $N_{1}$ and $N_{2}$) is given by
 \begin{align}
   \mathrm{C1'}: |h_{d}|+\sum_{n=1}^{N_{1}} |h_{1,n}||g_{1,n}|\geq \sqrt{2}\sum_{n=1}^{N_2}|h_{2,n}||g_{2,n}|.
 \end{align}
\end{lemma}
\begin{IEEEproof}
Please refer to Appendix \ref{appendix-partitioning-condition} for details.
\end{IEEEproof}
\vspace{-0.1cm}
\begin{rema}
Note that when the direct link is strong enough such that $|h_{d}|\geq \sqrt{2}\sum_{n=1}^{N_{2}}|h_{2,n}||g_{2,n}|$ holds with $N_{2}=N$, $\mathrm{C1'}$ can always hold under all possible pairs of $\{N_{1},N_{2}\}$ satisfying $N_{1}+N_{2}=N$. In this case, the constraint $\mathrm{C1'}$ can be relaxed without sacrificing the optimality.
\end{rema}
\vspace{-0.1cm}
With the transformed constraint $\mathrm{C1'}$, we next optimize the surface partitioning strategy of the two sub-surfaces, which can be expressed as
\begin{subequations} \nonumber
	\begin{align}
	\underline{\textbf{\text{P3:}}} \quad 
	&\max \limits_{N_{1},N_{2}}\ D_{\min}\triangleq\min\limits_{m \neq k}\left\{D_{m,k},m,k=1,\cdots,|\mathcal{A}_{x}|\right\}\\
	&\ \mbox{s.t.} \ \: \mathrm{C1'}, \mathrm{C3}.
	\end{align}
\end{subequations}

Although the optimal solution to \textbf{P3} can be obtained via a one-dimensional exhaustive search,
it lacks insights into the surface partitioning strategy. Hence, we consider the LoS scenario to obtain a closed-form solution and draw useful insights, which is a widely used assumption in the existing works, e.g., \cite{han2019large}. In this scenario, all the channel coefficients are composed of the large-scale path loss and the fixed phase shifts, where the path loss of the PTx-CRx link, the PTx-RIS link, and the RIS-CRx link are denoted by $\rho_{1}$, $\rho_{2}$, $\rho_{3}$, respectively.
For ease of derivation, we relax the integer constraints into the continuous space. When optimal $N_{1}$ and $N_{2}$ are obtained, we use the integer rounding technique to reconstruct the integer solution.

Based on the LoS channel assumption, the feasible condition $\mathrm{C1'}$ can be recast as
 \begin{align} \label{eq: C1;;}
    \mathrm{C1''}: N_{1}\geq \widetilde{N}_{1}=\begin{cases}(2-\sqrt{t^2+2})N
     &, 0\leq t\leq\sqrt{2}\\
     0 &,t\geq \sqrt{2}
     \end{cases}
 \end{align}
where $t\!=\!\sqrt{\frac{\rho_{1}}{\rho_{2}\rho_{3}N^2}}$ is defined as the channel strength ratio of the direct link to the reflected link.

Next, we only need to analyze the monotonicity of the objective function $D_{\min}$ with respect to (w.r.t.) the variable $N_{1}$, and find its maximum together with the conditions $\mathrm{C1''}$ and $\mathrm{C3}$. Overall, the solutions to \textbf{P3} are summarized in Theorem \ref{thm-elements-allocation}.
\begin{thm} \label{thm-elements-allocation}
The optimal surface partitioning strategies to \textbf{P3} under LoS channels are shown as
\begin{itemize}
     \item Case 1: $0\leq t\leq \sqrt{3}-\sqrt{2}$, the optimal partitioning strategy is given by 
\begin{small}
        \begin{align} \label{eq: optimal_partitioning_strategy}
        N_{1}^{\star}&=
        \left \lceil{\frac{(\sqrt{3}+1-\sqrt{2})-(\sqrt{2}+1-\sqrt{3})t}{2}N}\right \rceil, \\
        N_{2}^{\star}&=\left \lceil{\frac{\sqrt{2}+1-\sqrt{3}+(\sqrt{2}+1-\sqrt{3})t}{2}N}\right \rceil,
    \end{align}
\end{small}
    \item Case 2: $\sqrt{3}-\sqrt{2}\leq t \leq \sqrt{3}(\sqrt{2}-1)$, the optimal partitioning strategy is given by
    \begin{align} N_{1}^{\star}\!=\! \left \lceil{(2-\sqrt{t^2+2})N}\right\rceil, N_{2}^{\star}\!=\!N-\left \lceil{(2-\sqrt{t^2+2})N}\right\rceil
    \nonumber
    \end{align}
    \item Case 3: $\sqrt{3}(\sqrt{2}-1)\!\leq \!t \leq \frac{\sqrt{2}+\sqrt{6}}{2}$, the optimal partitioning strategy is the same as Case 1.
    \item Case 4: $t\geq \frac{\sqrt{2}+\sqrt{6}}{2}$, the optimal partitioning strategy is $N_{1}^{\star}=0$, $N_{2}^{\star}=N$.
\end{itemize}
\end{thm}
\begin{IEEEproof}
	Please refer to Appendix \ref{appendix-elements-allocation} for details.
\end{IEEEproof}

Theorem \ref{thm-elements-allocation} derives the optimal surface partitioning strategy under LoS channels with explicit physical meanings. The obtained strategy depends on the factor $t=\sqrt{\frac{\rho_{1}}{\rho_{2}\rho_{3}N^2}}$, which can be interpreted as the channel strength ratio of the direct link to the reflected link when the RIS phase shifts are optimized. It is observed that with the increase in this ratio, $N_{1}$ gradually decreases while $N_{2}$ increases. This can be explained as follows. For the case of small $t$ (e.g., direct link is relatively weak or even blocked due to obstacles), the RIS needs to allocate more reflecting elements for sub-surface \uppercase\expandafter{\romannumeral1} to assist the primary transmission. When $t$ becomes larger, the RIS can allocate fewer elements for the sub-surface \uppercase\expandafter{\romannumeral1}, while more elements can be allocated for the sub-surface \uppercase\expandafter{\romannumeral2} to enhance the performance of the secondary transmission. When $t$ is greater than a threshold (e.g., the direct link is very strong), the primary transmission does not need the dedicated help from the sub-surface \uppercase\expandafter{\romannumeral1}. In this case, we have $N_{1}^{\star}=0$, $N_{2}^{\star}=N$. In addition, although we just investigate the surface partitioning strategy in the LoS scenario, the solutions to this scenario also show a good performance in general channel models, e.g., Rician fading, which will be evaluated in the simulations of Sec. \ref{sec-simulation-results}.

In practice, the primary transmission usually has a higher service priority than the secondary transmission. In this case, we can add the constraint of performance requirement of the primary transmission to \textbf{P1}, which is formulated as a new problem \textbf{P-A}.
\begin{subequations} \label{eq:priority}
	\begin{align}
	\underline{\textbf{\text{P-A:}}}  
	&\max \limits_{\bm{\Phi}_{1},\bm{\Phi}_{2},N_{1},N_{2}}\!  D_{\min}\\
	&\quad  \quad  \mbox{s.t.} \ \: \mathrm{C1-C5}, \\ & \quad \quad \quad \ \ D_s \geq D_{s,0}, \label{eq: P-A-c}
	\end{align}
\end{subequations}
where $D_{s}\triangleq\min\limits_{m\neq k}\{D_{m,k},s_{m}\neq s_{k}\}$ denotes the minimum Euclidean distance of the primary signal; $D_{s,0}$ denotes the distance requirement of the constellation of the primary signal. In general, a larger $D_{s,0}$ means a higher BER requirement of the primary transmission. Similar to \textbf{P1}, \textbf{P-A} can be solved by decoupling it into two sub-problems. For the sub-problem of phase shifts optimization, the closed-form solution derived in Theorem \ref{thm-phase-shifts} can be applied. For the sub-problem of surface partitioning strategy optimization, we can obtain a
set of constraints when $D_s$ takes each element from the set $\{D_{m,k},s_{m}\neq s_{k}\}$, and then derive a set of feasible regions of $N_{1}$ and $N_{2}$. For example, if $D_s=2|h_d+h_{r,1}|$, by solving the inequality $D_s\geq D_{s,0}$, we can obtain a feasible region of $N_1$ as $N_1\geq \frac{D_{s,0}}{2\sqrt{\rho_2\rho_3}}-\sqrt{\frac{\rho_1}{\rho_2\rho_3}}$ under LoS channels. 
With the feasible regions of $N_{1}$ and $N_{2}$, the surface partitioning strategy to \textbf{P-A} can be obtained by following the procedures in Appendix \ref{appendix-elements-allocation}, for which the details are omitted here for brevity.

\vspace{-0.3cm}
\subsection{Complexity Analysis}
The complexity of the overall scheme is analyzed as follows. The problem \textbf{P1} is decoupled into two sub-problems, which are related to the optimization of phase shifts and the surface partitioning strategy. For the first sub-problem, the complexity of calculating the phase shifts of sub-surface I via Eq. \eqref{eq: phase-shifts-I} is $\mathcal{O}(N_{1})$, while the complexity of calculating the phase shifts of sub-surface II via Eq. \eqref{eq: phase-shifts-II} is $\mathcal{O}(N_{2})$. Besides, the complexity of determining the common phase shift of sub-surface II is $\mathcal{O}(\frac{1}{2}(M_{s}M_{c}-1)M_sM_c)$ via formula \eqref{eq:-common-phase}. Here, $M_s$ and $M_c$ denote the cardinalities of the constellation sets of the primary and secondary signals, respectively.
For the second sub-problem, the optimal surface partitioning strategy is determined via closed-form solutions, for which the complexity is independent of the number of reflecting elements and thus is $\mathcal{O}(1)$. In summary, the overall complexity of the proposed scheme is $\mathcal{O}(N+\frac{1}{2}(M_{s}M_{c}-1)M_sM_c)$ with $N=N_1+N_2$, which depends on the total number of reflecting elements and the cardinalities of the primary and secondary constellations.
\vspace{-0.4cm}
\section{Performance Analysis}\label{sec-performance-analysis}
\vspace{-0.1cm}
In the previous section, we have optimized the phase shifts and the surface partitioning strategy. However, it still remains unknown how the coupling effect of SR affects the assistance-transmission tradeoff and how much gain our proposed RIS partitioning scheme can achieve. 
To answer these two questions, in this section, we first derive the general BER expressions. After that, we analyze the coupling effect on this tradeoff and characterize the performance gain.
\subsection{BER Analysis}
According to the proposed two-step detector, the composite signal is decoded first, followed by the mapping from the decoded composite signal to the primary and secondary signals. Correspondingly, the BERs of the composite signal $x$, primary signal $s$, and secondary signal $c$ can be calculated as
\vspace{-0.2cm}

\begin{small}
    \begin{align}
    P_{x}&\!=\!\sum_{m=1}^{|\mathcal{A}_{x}|} \!\sum_{k=1,k\neq m}^{|\mathcal{A}_{x}|} \!P(\hat{x}\!=\!x_{k}\big | x\!=\! x_{m})P(x\!=\!x_{m})\frac{e(x_{m} \! \rightarrow \! x_{k})}{\log_{2}(|\mathcal{A}_{x}|)},  \label{eq: Px} 
    \\
    P_{s}&\!=\!\sum_{m=1}^{|\mathcal{A}_{x}|} \!\sum_{k=1,k\neq m}^{|\mathcal{A}_{x}|} \!P(\hat{x}\!=\!x_{k}\big | x\!= \!x_{m})P(x\!=\!x_{m})\frac{e(s_{m}\!\rightarrow \! s_{k})}{\log_{2}(|\mathcal{A}_{s}|)}, \label{eq: Ps} 
    \\
    P_{c}&\!=\!\sum_{m=1}^{|\mathcal{A}_{x}|} \!\sum_{k=1,k\neq m}^{|\mathcal{A}_{x}|} \!P(\hat{x}\!=\!x_{k}\big | x\!= \!x_{m})P(x\!=\!x_{m})\frac{e(c_{m}\!\rightarrow \!c_{k})}{\log_{2}(|\mathcal{A}_{c}|)}, \label{eq: Pc} 
\end{align}
\end{small}
where $e(x_{m}\rightarrow x_{k})$, $e(x_{m}\rightarrow x_{k})$, and $e(x_{m}\rightarrow x_{k})$ denote the Hamming distances between the constellations $x_{m}$ and $x_{k}$, $s_{m}$ and $s_{k}$, $c_{m}$ and $c_{k}$, respectively; $P(x=x_{m})=P(s= s_{m})P(c=c_{m})$ denotes the probability when $x_{m}$ is transmitted, which is the product of transmitted probabilities of $s_{m}$ and $c_{m}$; $P(\hat{x}=x_{k}\big | x= x_{m})$ denotes the conditional probability of making a decision of $x_{k}$ when $x_{m}$ is transmitted. In general, it is difficult to calculate the conditional probability due to the irregular decision regions. Fortunately, it is reported in \cite{proakis2001digital} that pairwise error probability can serve as an upper-bound for $P(\hat{x}=x_{k}\big | x= x_{m})$, given by
\begin{align}\label{eq: BER-approximation}
P(\hat{x}=x_{k}\big | x= x_{m})&\leq P(x_{m}\rightarrow x_{k}) \nonumber \\&=P(|y-\sqrt{p_{t}}x_{m}|^2> |y-\sqrt{p_{t}}x_{k}|^2) \nonumber \\ &
=\mathcal{Q}(\mu D_{m,k}),
\end{align}
where $\mu=\sqrt{\frac{p_{t}}{2\sigma^{2}}}$, $D_{m,k}=|x_{m}-x_{k}|$. This bound is tight, especially for the high SNR regime~\cite{proakis2001digital}.

By substituting this upper bound into \eqref{eq: Px}, \eqref{eq: Ps}, and \eqref{eq: Pc}, it is observed that the BERs of $x$, $s$, and $c$ are coupled together, which depend not only on the Euclidean distances of symbol $x$ but also on the bit-mapping rule of symbol $x$. Moreover, since the $\mathcal{Q}$-function is a monotonically decreasing function, the minimum Euclidean distance term has a dominant effect on the BER performance. Thus, we will analyze the relationship between the minimum Euclidean distance and the corresponding bit-mapping rule to obtain the BER approximations. 

\begin{figure}[t]  
	\centering  
	\captionsetup{font={scriptsize}}
	\setlength{\abovecaptionskip}{-0.1cm}
	\includegraphics[width=2.2in]{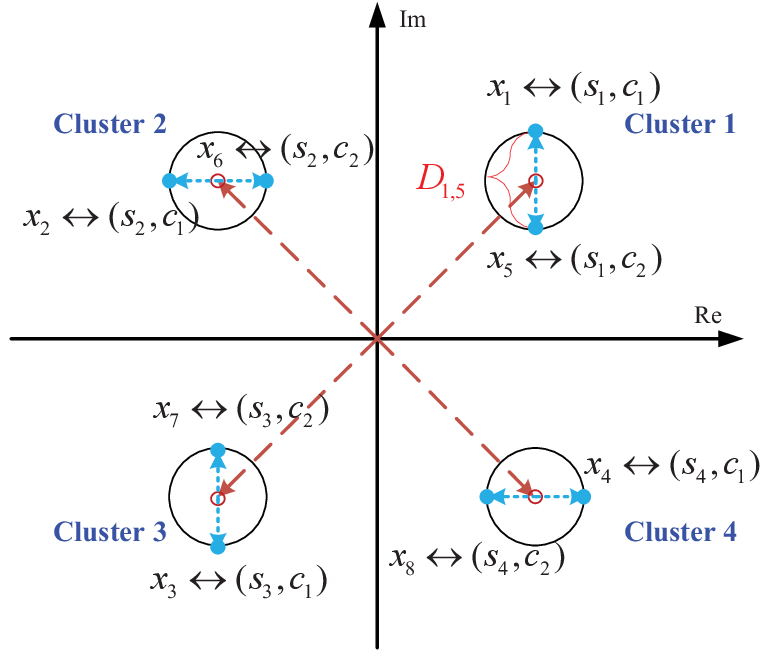}  
	\caption{Constellations of the composite signal $x=(h_{d}+h_{r,1}+h_{r,2}c)s$ when the primary and secondary signals (i.e., $s$ and $c$) adopt QPSK and BPSK, respectively. The red dashed lines denote $h_{d}+h_{r,1}$, while the blue dotted lines denote $h_{r,2}$. The blue dots denote the composite signal, and the composite signal $x$ comprising the same $s$ but different $c$ form a cluster (e.g., see Clusters $1$-$4$).}
	\label{fig:signal-space-composite-signal}  
\end{figure}

First, according to the optimized phase shifts in Theorem \ref{thm-phase-shifts}, we plot the constellations of the composite signal in Fig. \ref{fig:signal-space-composite-signal}, where the mapping rule between $x$ and $\{s,c\}$, as well as  Euclidean distances of the composite signal are summarized in Table \ref{table:Mapping-Euclidean-distance}. Specifically, all the Euclidean distances can be categorized into three types, described as follows
\begin{table*}[t]
\centering
\caption{Mapping Rule and Euclidean Distances of the Composite Signal.}
\begin{tabular}{c||c|c||c|c} 
\hline
Mapping rule                        & Index                   & Euclidean distance                                & Index                   & Euclidean distance                                                              \\ 
\hline
$x_{1}\leftrightarrow(s_{1},c_{1})$ & $D_{1,2}=|x_{1}-x_{2}|$ & $\sqrt{2}|h_{d}+h_{r,1}+h_{r,2}|$                 & $D_{5,6}=|x_{5}-x_{6}|$ & $\sqrt{2}|h_{d}+h_{r,1}-h_{r,2}|$                                               \\ 
\hline
$x_{2}\leftrightarrow(s_{2},c_{1})$ & $D_{1,3}=|x_{1}-x_{3}|$ & $2|h_{d}+h_{r,1}+h_{r,2}|$                        & $D_{5,7}=|x_{5}-x_{7}|$ & $2|h_{d}+h_{r,1}-h_{r,2}|$                                                      \\ 
\hline
$x_{3}\leftrightarrow(s_{3},c_{1})$ & $D_{1,4}=|x_{1}-x_{4}|$ & $\sqrt{2}|h_{d}+h_{r,1}+h_{r,2}|$                 & $D_{5,8}=|x_{5}-x_{8}|$ & $\sqrt{2}|h_{d}+h_{r,1}-h_{r,2}|$                                               \\ 
\hline
$x_{4}\leftrightarrow(s_{4},c_{1})$ & $D_{1,5}=|x_{1}-x_{5}|$ & $2|h_{r,2}|$                                      & $D_{5,1}=|x_{5}-x_{1}|$ & $2|h_{r,2}|$                                                                    \\ 
\hline
$x_{5}\leftrightarrow(s_{1},c_{2})$ & $D_{1,6}=|x_{1}-x_{6}|$ & $\sqrt{2}|h_{d}+h_{r,1}+h_{r,2}e^{\jmath\pi/2}|$  & $D_{5,2}=|x_{5}-x_{2}|$ & $\sqrt{2}|h_{d}+h_{r,1}+h_{r,2}e^{-\jmath\pi/2}|$                               \\ 
\hline
$x_{6}\leftrightarrow(s_{2},c_{2})$ & $D_{1,7}=|x_{1}-x_{7}|$ & $2|h_{d}+h_{r,1}|$                                & $D_{5,3}=|x_{5}-x_{3}|$ & $2|h_{d}+h_{r,1}|$                                                              \\ 
\hline
$x_{7}\leftrightarrow(s_{3},c_{2})$ & $D_{1,8}=|x_{1}-x_{8}|$ & $\sqrt{2}|h_{d}+h_{r,1}+h_{r,2}e^{-\jmath\pi/2}|$ & $D_{5,4}=|x_{5}-x_{4}|$ & $\sqrt{2}|h_{d}+h_{r,1}+h_{r,2}e^{\jmath\pi/2}|$                                \\ 
\hline
$x_{8}\leftrightarrow(s_{4},c_{2})$ & \multicolumn{4}{c}{Note:~$h_{r,1}=\sum_{n=1}^{N_{1}}|h_{1,n}||g_{1,n}|e^{\jmath \angle(h_{d})},~ h_{r,2}=\sum_{n=1}^{N_{2}}|h_{2,n}||g_{2,n}|e^{\jmath (\angle(h_{d})+\frac{\pi}{4})}$ according to Theorem \ref{thm-phase-shifts}.}  \\
\hline
\multicolumn{5}{c} {Primary signal: $s_{1}=\frac{1+\jmath}{\sqrt{2}}$, $s_{2}=\frac{-1+\jmath}{\sqrt{2}}$, $s_{3}=\frac{-1-\jmath}{\sqrt{2}}$, $s_{4}=\frac{1-\jmath}{\sqrt{2}}$;
Secondary signal: $c_{1}=1$, $c_{2}=-1$}  \\
\hline
\end{tabular}\label{table:Mapping-Euclidean-distance}
\end{table*}

\begin{itemize}
    \item \textbf{Intra-cluster Distance}: The Euclidean distance between the constellations in the same cluster (e.g., Cluster 1), given by
    \begin{align}
        D^{I}=D_{1,5}=2|h_{r,2}|.
    \end{align}
    \item \textbf{Inter-adjacent-cluster Distance}: The Euclidean distance between the constellations in the adjacent clusters (e.g., Cluster 1 and Cluster 4), given by
    \begin{align}
        D^{II}&=\min\{D_{1,8},D_{1,4},D_{5,4},D_{5,8}\}
        \nonumber
        \\
        &= \sqrt{2}|h_{d}+h_{r,1}+h_{r,2}e^{\jmath \pi/2}|.
    \end{align}
    \item \textbf{Inter-non-adjacent-cluster Distance}: The Euclidean distance between the constellations in the non-adjacent clusters (e.g., Cluster 1 and Cluster 3), given by
    \begin{align}
        D^{III}\!&=\!\min\{D_{1,3}, D_{1,7}, D_{5,3}, D_{5,7}\} \nonumber
        \\
        \!&=\! 
        \begin{cases}
        2|h_{d}\!+\!h_{r,1}|, \ \text{if} \ |h_{r,2}|^2\geq 2\Re\{(h_{d}\!+\!h_{r,1})h_{r,2}^{H}\}
        \\
        2|h_{d}+h_{r,1}-h_{r,2}|, \ \text{otherwise}
        \end{cases} \nonumber
    \end{align}
\end{itemize}

Since the minimum Euclidean distance $D_{\min}$ has a dominant impact on BER, it is reported in~\cite{proakis2001digital} that the BER can be approximated as $\mathcal{Q}(\mu D_{\min})$, where $\mu=\sqrt{\frac{p_{t}}{2\sigma^{2}}}$. In our work, the minimum Euclidean distance of the composite signal is the minimum of the above three types of distances, i.e., $D_{\min}=\min\{D^{I},D^{II},D^{III}\}$
. With the BER approximation, we can analyze the coupling effect in the following.

\vspace{-0.4cm}
\subsection{Coupling Effect Analysis} \label{sec: coupling effect}
Based on the BER expressions derived above, we analyze the coupling effect from a composite signal perspective in the absence and presence of the direct link, respectively.
\subsubsection{In the absence of the direct link}
\begin{figure}[!t]
	\centering
	\setlength{\abovecaptionskip}{-0.08cm}
	\captionsetup{font={scriptsize}}
	\subfigcapskip = -8pt
	\subfigure[] {
		\label{fig:fig-MED-wo-Direct-link}    
		\includegraphics[width=0.35\textwidth]{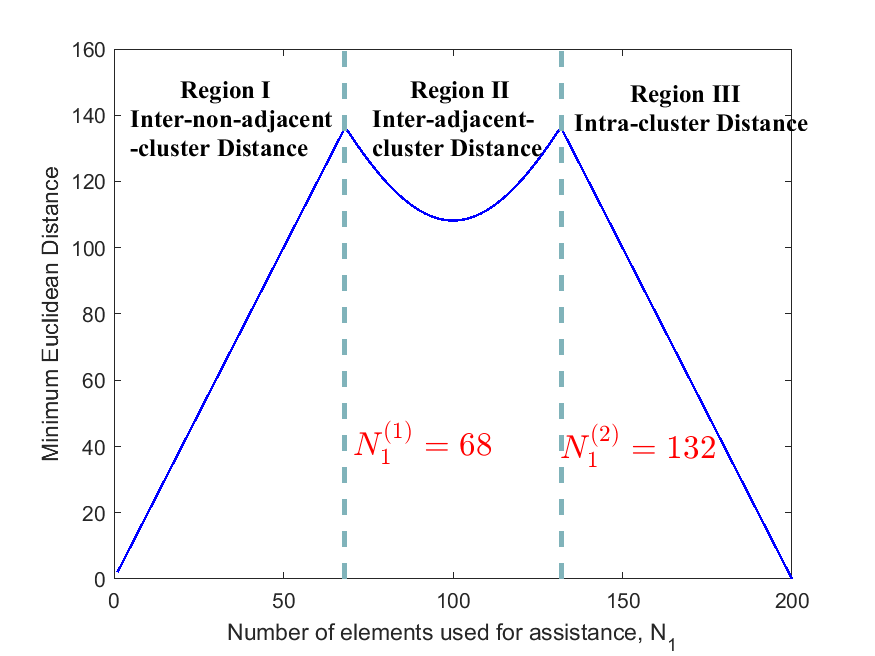}
		\captionsetup{font={scriptsize}}
	}
	\subfigure[]{
		\label{fig:fig-EBR-wo-Direct-link} 
		\includegraphics[width=0.35\textwidth]{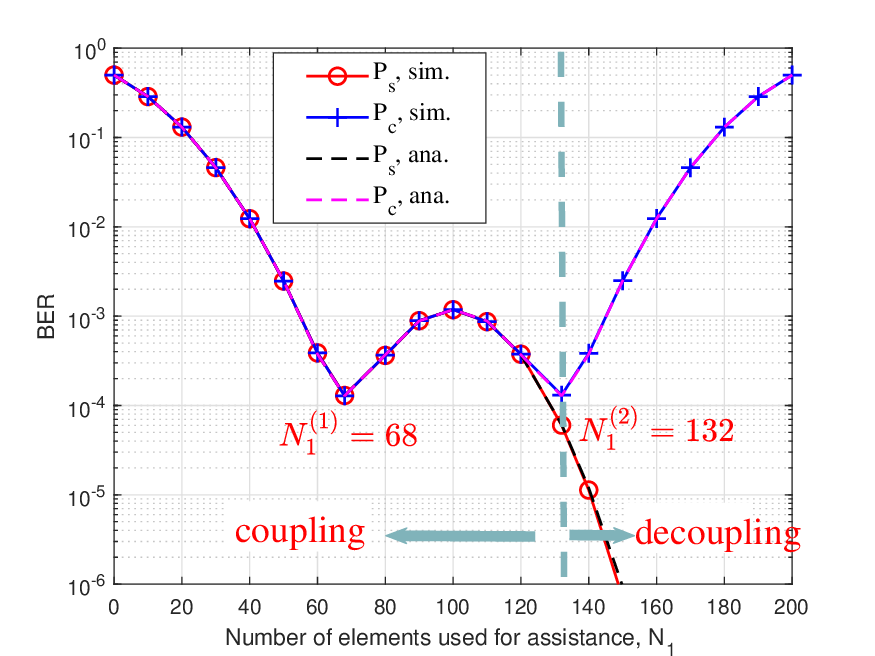}
	}
	\caption{ In the absence of the direct link, where $N_{1}^{(1)}=\left \lceil\frac{\sqrt{2}+1-\sqrt{3}}{2}N \right\rceil$, $N_{1}^{(2)}=\left \lceil\frac{\sqrt{3}+1-\sqrt{2}}{2}N\right\rceil$ by assuming $N=200$. (a) The minimum Euclidean distance vs. $N_{1}$; (b) The BERs of the primary and secondary signal vs. $N_{1}$.}\label{fig: without-direct-link}
\end{figure}

We assume the direct link is blocked where $h_{d}=0$. Recall that our optimized surface partitioning strategy is obtained under LoS channels, so we plot the minimum Euclidean distance $D_{\min}$ versus the number of elements used for assistance in Fig. \ref{fig:fig-MED-wo-Direct-link} under LoS channels to perform our analysis.
As we can see, $D_{\min}$ is shown as a piece-wise function, which is discussed in the following three regions.

\emph{Region I $\left(0\leq N_{1} < \left\lceil\frac{\sqrt{2}+1-\sqrt{3}}{2}N\right\rceil\right)$}: The inter-non-adjacent-cluster distance $D^{III}=D_{1,7}=2|h_{d}+h_{r,1}|$ is the minimum, which corresponds to the distance between the constellations $x_{1}$ and $x_{7}$. This means that when symbol $x_{1}$ (i.e., $s_{1},c_{1}$) is transmitted, the most probable error is to select symbol $x_{7}$ (i.e., $s_{3},c_{2}$) as the true one. In this case, both the primary and secondary signals will be decoded wrongly. Therefore, both the BERs of $s$ and $c$ would be limited by this minimum distance, and this is the explanation of the coupling effect from a composite signal perspective. Accordingly, using \eqref{eq: BER-approximation}, the BERs are given by 
\begin{align}
    P_{s}\approx P_{c} \approx \mathcal{Q}(2\mu |h_{r,1}|).
\end{align}

\emph{Region II $\left(\left \lceil\frac{\sqrt{2}+1-\sqrt{3}}{2}N \right\rceil< N_{1} < \left \lceil\frac{\sqrt{3}+1-\sqrt{2}}{2}N\right\rceil\right)$}: The inter-adjacent-cluster distance $D^{II}=D_{5,4}=\sqrt{2}|h_{d}+h_{r,1}+h_{r,2}e^{\jmath \pi/2}|$ becomes the minimum, which corresponds to the distance between the constellations $x_{5}$ and $x_{4}$. This means that when symbol $x_{5}$ (i.e., $s_{1},c_{2}$) is transmitted, the most probable error occurs when selecting symbol $x_{4}$ (i.e., $s_{4},c_{1}$) as the true one. In this case, both the primary and secondary signals will be decoded wrongly. Thus, both the BERs of $s$ and $c$ would be limited by this minimum distance, resulting in the coupling, given by
\begin{align}
    P_{s}\approx P_{c} \approx \mathcal{Q}(\sqrt{2}\mu|h_{r,1}+h_{r,2}e^{\jmath \pi/2}|).
\end{align}

\emph{Region III $\left( \left\lceil\frac{\sqrt{3}+1-\sqrt{2}}{2}N\right\rceil < N_{1} \leq N\right)$}: The intra-cluster distance $D^{I}=D_{1,5}=2|h_{r,2}|$ becomes the minimum, which corresponds to the distance between the constellations $x_{1}$ and $x_{5}$. This means that when symbol $x_{1}$ (i.e., $s_{1},c_{1}$) is transmitted, the most probable error is to select symbol $x_{5}$ (i.e., $s_{1},c_{2}$) as the true one. Notably, in this case, only the secondary signal $c$ will be decoded wrongly. Thus, the BERs of the primary and secondary signal will be limited by intra-cluster distance and the inter-adjacent-cluster distance, respectively, yielding different BER limits,
\begin{align}
    P_{s}\approx\mathcal{Q}(\sqrt{2}\mu|h_{r,1}+h_{r,2}e^{\jmath \pi/2}|), \ P_{c}\approx\mathcal{Q}(2\mu|h_{r,2}|).
\end{align}

To evaluate the above analysis, we plot the BERs of the primary and secondary signals in Fig. \ref{fig:fig-EBR-wo-Direct-link} under LoS channels, from which we see the theoretical results are consistent with the simulation results and summarize the relationships among the assistance-transmission tradeoff, the coupling effect, and the surface partitioning strategies as follows.
\begin{itemize}
    \item When $0\leq N_{1}\leq\left \lceil\frac{\sqrt{3}+1-\sqrt{2}}{2}N\right\rceil$, the coupling effect dominates the BERs of the primary and secondary signals, which yields the same BER limits. In this case, the phenomenon of assistance-transmission tradeoff cannot be observed.
    \item When $N_{1}\geq \left \lceil\frac{\sqrt{3}+1-\sqrt{2}}{2}N\right\rceil$, increasing $N_{1}$ means using more elements for assistance, thereby enhancing the performance of primary signal (i.e., $P_{s}$ decreases). However, in this case, fewer elements are used for information transmission, thereby deteriorating the performance of the secondary signal (i.e., $P_{c}$ increases). The phenomenon of assistance-transmission tradeoff is thus observed. Besides, $N_{1}^{(2)}=\left\lceil\frac{\sqrt{3}+1-\sqrt{2}}{2}N\right\rceil$ can be regarded as the transition point between the coupling and decoupling states, which achieves the best assistance-transmission tradeoff. Note that from the perspective of the composite signal, the decoupling state means the performance of the primary and secondary signals are limited by different Euclidean distance terms.
\end{itemize}
Moreover, from Fig. \ref{fig: without-direct-link} (a), we notice that the minimum Euclidean distance of the composite signal is symmetrical w.r.t. the line $N_{1}=\frac{N}{2}$ and attains its maximum for either $N_{1}=68$ or $N_{1}=132$. As for the two points, from Fig. \ref{fig: without-direct-link} (b), both $P_{s}$ and $P_{c}$ could achieve similar BER performance, which motivates us to investigate this phenomenon in Sec. \ref{sec-commutative-property}.
\subsubsection{In the presence of the direct link}
In this scenario, the $D_{\min}$-vs-$N_{1}$ curve will be left shifted compared to the case where the direct link is blocked, and the transition point becomes $N_{1}^{(2)}=\left \lceil{\frac{(\sqrt{3}+1-\sqrt{2})-(\sqrt{2}+1-\sqrt{3})t}{2}N}\right \rceil$ (see Theorem \ref{thm-elements-allocation}), where $t=\sqrt{\frac{\rho_{1}}{\rho_{2}\rho_{3}N^2}}$ is defined in \eqref{eq: C1;;}. Likewise, similar analysis can be conducted according to the case of a blocked direct link.

\vspace{-0.4cm}
\subsection{Performance Gain over Conventional RIS Design for SR}\label{sec-performance-gain}
\vspace{-0.1cm}
In this subsection, we characterize the performance gain brought by our proposed RIS partitioning scheme. The conventional RIS design is taken as the benchmark~\cite{zhou2022cooperative,hua2021uav,zhang2021reconfigurable,hua2021novel,hu2020reconfigurable}, which assigns all the
reflecting elements to information transmission (i.e., $N_{1}=0, N_{2}=N$). Next, we take the scenario without the direct link as an example to characterize the performance gain.
\subsubsection{BER Performance under Conventional RIS Design}
With the conventional RIS design scheme, the received signal at the CRx in the absence of the direct link is written as 
\begin{align}
    y=\sqrt{p_{t}}\bm{h}_{2}^{H}\bm{\Phi}_{2}\bm{g}_{2}s c+z, \label{eq: received-signal-wo-D}
\end{align}
where $\bm{h}_{2}$ and $\bm{g}_{2}$ are both $N$-dimensional vectors.  However, it is observed that the CRx encounters an ambiguity problem when performing joint decoding. Taking $\mathcal{A}_{s}\!=\! \mathcal{A}_{c}\!=\!\{1,-1\}$ as an example, both $\{\hat{s}_{1}(n)\!=\!1,
\hat{c}_{1}(n)\!=\!1\}$ and $\{ \hat{s}_{2}(n)\!=\!-1, \hat{c}_{2}(n)\!=\!-1\}$ are the optimal estimators to \eqref{eq: received-signal-wo-D} due to $\hat{s}_{1}(n)\hat{c}_{1}(n)\!=\!\hat{s}_{2}(n)\hat{c}_{2}(n)$. Thus, the ambiguity problem leads to the failure of joint decoding, namely, $P_{s}^{\mathrm{con}}=P_{c}^{\mathrm{con}}=0.5$.

\subsubsection{BER Performance under Proposed RIS Design}
When the proposed RIS design is adopted, the received signal at the CRx in the absence of the direct link is given by
\begin{align}
    y=\sqrt{p_{t}}\bm{h}_{1}^{H}\bm{\Phi}_{1}\bm{g}_{1}s+\sqrt{p_{t}}\bm{h}_{2}^{H}\bm{\Phi}_{2}\bm{g}_{2}s c+z, \label{eq: received-signal-w-D}
\end{align}
where $\bm{h}_{1}$ and $\bm{g}_{1}$ are both $N_{1}$-dimensional vectors, and $\bm{h}_{2}$ and $\bm{g}_{2}$ are both $N_{2}$-dimensional vectors. By adopting the proposed scheme, the ambiguity problem can be addressed since the reflected link via the sub-surface \uppercase\expandafter{\romannumeral1} can be regarded as the virtual direct link. Then, under the optimized surface partitioning strategy, i.e., the intersection of \emph{Region II} and \emph{Region III} in Sec. \ref{sec: coupling effect}, the BERs of the primary and secondary signals under LoS channels are given by
\begin{align}
    P_{s}^{\mathrm{pro}}&=\mathcal{Q}(\sqrt{2}\mu|h_{r,1}\!+\!h_{r,2}e^{\jmath \pi/2}|),\\
    P_{c}^{\mathrm{pro}}&=\mathcal{Q}(\sqrt{2}\mu|h_{r,1}\!+\!h_{r,2}e^{\jmath \pi/2}|)+\mathcal{Q}(2\mu|h_{r,2}|), 
\end{align}
where $\mu=\sqrt{\frac{p_{t}}{2\sigma^{2}}}$. By subtracting BERs $P_{s}^{\mathrm{pro}}$ and $P_{c}^{\mathrm{pro}}$ from BERs $P_{s}^{\mathrm{con}}$ and $P_{c}^{\mathrm{pro}}$, respectively, we obtain the performance gain of our proposed RIS design. To gain more insights, we focus on the region of large $N$ and derive a closed-form expression shown in Theorem \ref{thm: performance-gain}.
\vspace{-0.3cm}
\begin{thm}
\label{thm: performance-gain}
For a large $N$ and a blocked direct link, the performance gain of our proposed RIS design over the existing counterpart under LoS channels is given by
\begin{align}
    P_{s}^{\mathrm{gain}}\!=\! P_{s}^{\mathrm{con}}-P_{s}^{\mathrm{pro}}&=0.5-\mathcal{Q}\left(\sqrt{\alpha_{1} \gamma_{b}N^{2}}\right)
    \nonumber \\ &
    \overset{a}{\geq} 0.5- 0.5e^{-\frac{\alpha_{1}}{2}\gamma_{b}N^{2}},
\end{align}
\begin{align} 
P_{c}^{\mathrm{gain}}&=P_{c}^{\mathrm{con}}-P_{c}^{\mathrm{pro}}
    \nonumber \\ &=0.5-\mathcal{Q}\left(\sqrt{\alpha_{1}\gamma_{b}N^{2}}\right)\!-\!\mathcal{Q}\left(\sqrt{2\alpha_{2}\gamma_{b}N^2}\right) \nonumber\\
    &\!\overset{b}{\geq}0.5\!-\!0.5e^{-\frac{\alpha_{1}}{2}\gamma_{b}N^{2}}\!-\!0.5e^{-\alpha_{2}\gamma_{b}N^2},
\end{align}
where $\alpha_{1}=\zeta_{1}^{2}+\zeta_{2}^{2}-\sqrt{2}\zeta_{1}\zeta_{2}$ and  $\alpha_{2}=\zeta_{2}^{2}$ with $\zeta_{1}=\frac{\sqrt{3}+1-\sqrt{2}}{2}$, $\zeta_{2}= \frac{\sqrt{2}+1-\sqrt{3}}{2}$ obtained from the Case 1 in Theorem \ref{thm-elements-allocation}; $\gamma_{b}=\frac{p_{t}\rho_{2}\rho_{3}}{\sigma^{2}}$ is defined as the SNR of the reflected link; ``$a$'' and ``$b$'' hold due to the fact $\mathcal{Q}(\sqrt{\mathrm{SNR}})\leq \frac{1}{2}e^{-\mathrm{SNR}/2}$.
\end{thm}
\begin{IEEEproof}
	Please refer to Appendix \ref{appendix-performance-gain} for details.
\end{IEEEproof}

Theorem \ref{thm: performance-gain} shows that the gain is determined by the reflected link SNR $\gamma_{b}$, the optimized surface partitioning strategy, and the total number of reflecting elements $N$. Besides, we observe that the BERs under the proposed RIS design decay with the exponential of $N^{2}$, while the BERs under the conventional scheme are always $0.5$ regardless of $N$. Moreover, when the direct link is blocked, the RIS-assisted SR system under the conventional RIS design fails to work. However, with our proposed scheme, the system can still work normally and support the primary and secondary transmissions simultaneously thanks to the existence of sub-surface \uppercase\expandafter{\romannumeral1}.
As for the scenario with the direct link, the performance gain can also be analyzed in a similar way by using the optimal surface partitioning strategy obtained in Theorem \ref{thm-elements-allocation} and the BER analysis in Sec. \ref{sec: coupling effect}, which will be evaluated in the simulation results.

\vspace{-0.4cm}
\section{Commutative Property  of Assistance-Transmission Dual Capabilities} \label{sec-commutative-property}
Based on the above performance analysis, we observe that the $D_{\min}$-vs-$N_{1}$ curve exhibits a strictly symmetric property w.r.t. the line $N_{1}=N/2$ from Fig. \ref{fig:fig-MED-wo-Direct-link}.
Inspired by this observation, we delve into the physical meaning behind it and find an interesting property, which is named as the commutative property of assistance-transmission dual capabilities in this paper. 

Next, we will take the scenario where the direct link is blocked as an example to analyze this commutative property. To begin with, we present an important Lemma regarding the composite signal to help understand the commutative property, shown as follows.
\begin{lemma}
\label{lem: commutative-property}
Given any arbitrary two complex numbers $a=|a|e^{\jmath\theta_{a}}$ and $b=|b|e^{\jmath\theta_{b}}$, we let $f_{1}(c)= |a|e^{\jmath\theta_{a}}+c|b|e^{\jmath\theta_{b}}$, $f_{2}(c)=|b|e^{\jmath\theta_{a}}+c|a|e^{\jmath\theta_{b}}$, where $c\in \mathcal{A}_{c}=\{1,-1\}$ denotes the secondary signal. Then, the following equalities hold
\begin{small}
    \begin{align}
    &\quad \quad \quad \quad \quad \quad \quad |f_{1}(c)|=|f_{2}(c)|, \forall c \in \mathcal{A}_{c}, \label{eq: amplitude}\\
   & \underbrace{\angle\left(f_{1}(1)\right)-\angle\left(f_{1}(-1)\right)}_{\text{Phase difference for different $c$}} + \underbrace{\angle\left(f_{2}(1)\right)-\angle\left(f_{2}(-1)\right)}_{\text{Phase difference for different $c$}}= \pi. \label{eq: phase}
\end{align}
\end{small}
\end{lemma}
\begin{IEEEproof}
	It can be proved by expanding the forms of the amplitude and the phase of $f_{1}(c)$ and $f_{2}(c)$, respectively. The details are omitted here for brevity.
\end{IEEEproof}

In Lemma \ref{lem: commutative-property}, the complex numbers $a=|a|e^{\jmath\theta_{a}}$ and $b=|b|e^{\jmath\theta_{b}}$ can be viewed as the equivalent reflected links via the sub-surfaces \uppercase\expandafter{\romannumeral1} and sub-surface \uppercase\expandafter{\romannumeral2}, respectively; and $c$ can be regarded as the secondary signal. Then, $f_{1}(c)$ and $f_{2}(c)$ can be viewed as the two types of composite signals when we exchange the number of elements allocated for these two sub-surfaces. Based on \eqref{eq: amplitude} and \eqref{eq: phase}, Lemma \ref{lem: commutative-property} tells us that the exchange of the number of elements between assistance and transmission does not change the amplitude of the composite signal but only change the phase rotation of the composite signal under different $c$, thereby yielding symmetric constellations of the composite signal.
\begin{figure}[!t]
		\centering
		\captionsetup{font={scriptsize}}
		\subfigure[] {
			\label{fig:assistance-geq-transmission-wo-D}
			\includegraphics[width=0.22\textwidth]{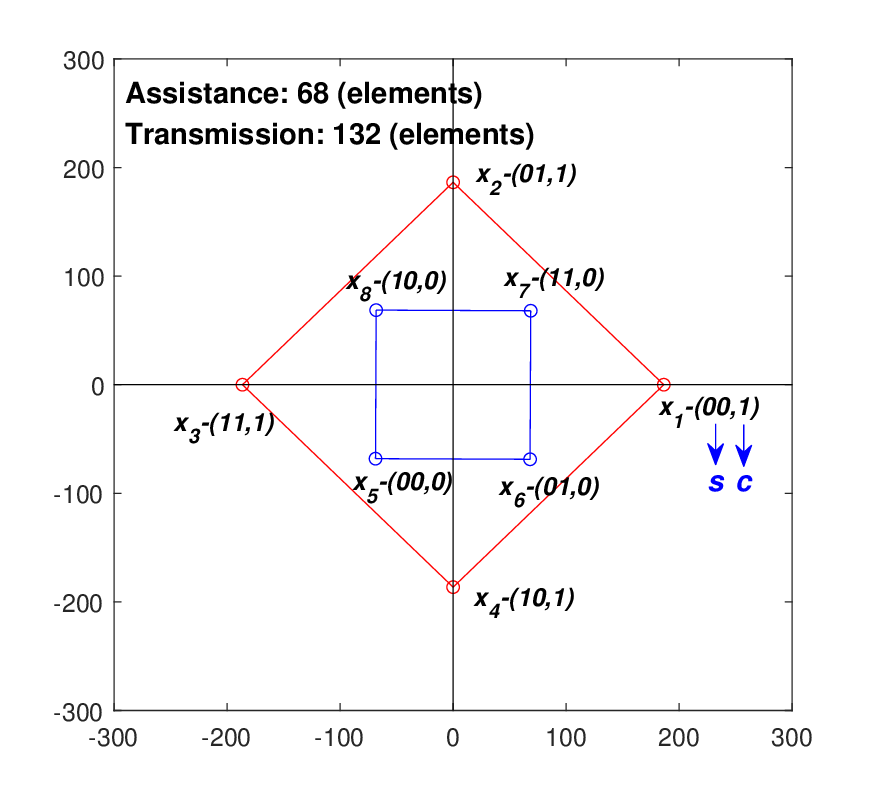}
		}
		\subfigure[]{
			\label{fig:assistance-leq-transmission-wo-D}
			\includegraphics[width=0.22\textwidth]{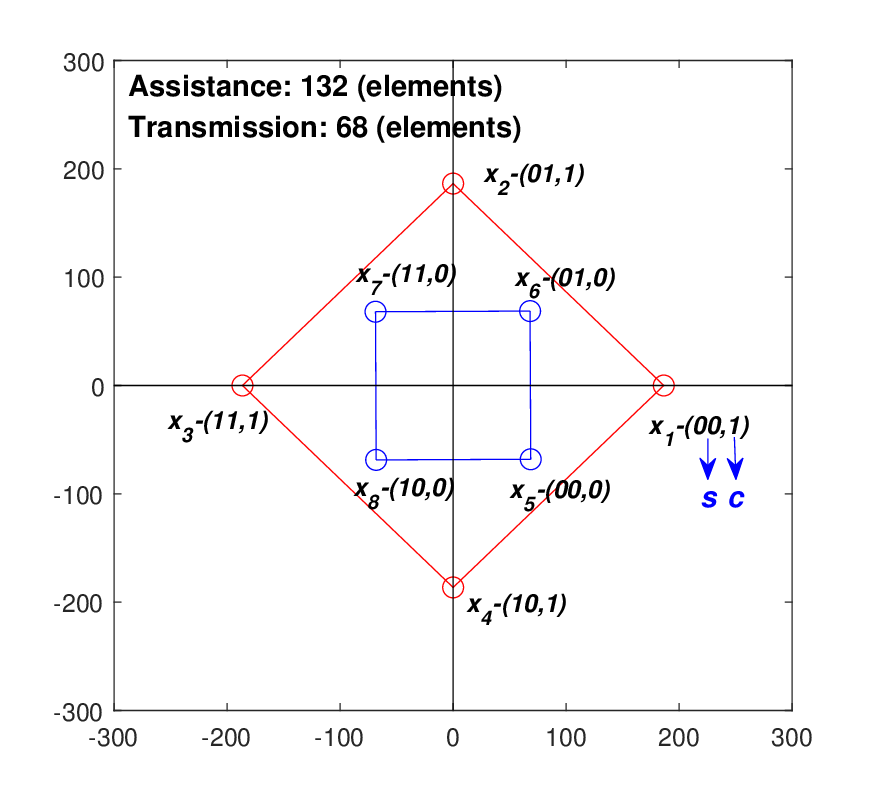}
		}
	\vspace{-0.3cm}
		\caption{Assume the total number of reflecting elements $N=200$. (a) \emph{Scheme I}:  the constellations of the composite signal under $N_{1}=68$, $N_{2}=132$; (b) \emph{Scheme II}:  the constellations of the composite signal under $N_{1}=132$, $N_{2}=68$ .}
	\end{figure}


For ease of understanding, we take the two points (i.e., $N_{1}^{(1)}=\left \lceil\frac{\sqrt{2}+1-\sqrt{3}}{2}N \right\rceil$, $N_{1}^{(2)}=\left \lceil\frac{\sqrt{3}+1-\sqrt{2}}{2}N\right\rceil$) in Fig. \ref{fig:fig-MED-wo-Direct-link} as an example and plot the constellations of the composite signal under these two schemes in Fig. \ref{fig:assistance-geq-transmission-wo-D} and Fig. \ref{fig:assistance-leq-transmission-wo-D} by assuming $N=200$. It is observed that the constellations under these two schemes are strictly symmetrical w.r.t. the $y$-axis but possess different bit-mapping relationships. This observation explains why the BERs of $P_{s}$ and $P_{c}$ under $N_{1}=68$ and $N_{1}=132$ in Fig. \ref{fig: without-direct-link}(b) could achieve the similar performance.
Such findings reveal a commutative property of
assistance-transmission dual capabilities, defined as

\textbf{Definition 1}. \emph{\textbf{(Commutative Property)}} \emph{When the direct link is blocked, exchanging the number of reflecting elements between assistance and transmission does not change the SER performance of the composite signal thanks to the symmetrical constellations, but only slightly changes its BER performance due to different bit-mapping rules.}

As for the performance of the primary and secondary signals, the commutative property only holds in case the number of elements allocated for assistance belongs to the interval $N_{1}^{(1)}=\left \lceil\frac{\sqrt{2}+1-\sqrt{3}}{2}N \right\rceil\leq N_{1}\leq N_{1}^{(2)}=\left \lceil\frac{\sqrt{3}+1-\sqrt{2}}{2}N\right\rceil$, which can be readily verified from Fig. \ref{fig: without-direct-link} (b). 

Rigorously speaking, scheme II is better than scheme I due to the impact of phase rotation, which will be evaluated in Fig. \ref{fig:fig_tradeoff_wo_D}.
This is because although the two surface partitioning strategies of scheme I and scheme II could both maximize the minimum Euclidean distance of the composite signal, the two surface partitioning strategies corresponds to two different phase rotations, which leads to different bit-mapping rules. In this regard, scheme II outperforms scheme I since the phase rotation of its partitioning strategy leads to a better bit-mapping rule of the composite signal, as discussed in Sec. \ref{sec:Hamming-distance}.
\vspace{-0.2cm}
\begin{rema}\label{rema: commutative}
The commutative property can be readily extended to the scenario with the direct link, which will be detailed and evaluated in Fig. \ref{fig: fig_tradeoff} (b) of Sec. \ref{sec-simulation-results}.
\end{rema}
\vspace{-0.3cm}
\section{RIS Design for Higher-Order Modulation Constellations} \label{sec:-extension}
In the previous sections, we use QPSK and BPSK for primary and secondary signals as an example to facilitate our design and draw useful insights. In this section, we extend our design and analysis to the cases of higher-order modulation constellations, e.g., QAM primary signal and PSK secondary signal.

In this case, the criteria of the minimum Euclidean distance and the Hamming distance are still applicable, while the constraints of the Hamming distance depend on the specific modulation constellations. To build a general optimization framework, we can first solve the problem of maximizing the minimum Euclidean distance of the composite signal, by optimizing the phase shifts and surface partitioning strategy. 
Specifically, the Euclidean distance of the composite signal can be classified into the following three types.
\begin{align} 
	\begin{cases} D_{1}\!=\!
	|\bm{\phi}_{2}^{H}\bm{f}_{2}|^2|c_{m}-c_{k}|^2, \quad  \  s_{m}=s_{k},c_{m}\neq c_{k}\\
	D_{2}\!=\!|h_{d}\!+\!\bm{\phi}_{1}^{H}\bm{f}_{1}|^2|s_{m}\!-\!s_{k}|^2, s_{m}c_{m}\!=\!s_{k}c_{k},s_{m}\!\neq\! s_{k}, c_{m}\!\neq\! c_{k} \\
	D_{3}\!=\!\left|h_{d}\!+\!\bm{\phi}_{1}^{H}\bm{f}_{1}\!+\!\bm{\phi}_{2}^{H}\bm{f}_{2}\frac{s_{m}\!-\!s_{k}\frac{c_k}{c_m}}{s_{m}\!-\!s_{k}}c_m\right|^2\!|s_{m}\!-\!s_{k}|^2,  \mathrm{otherwise}.
\end{cases} \nonumber
\end{align}

Similar to our former design methodology, the problem of maximizing the minimum of $\{D_1,D_2,D_3\}$ can be divided into two sub-problems, one related to the phase shifts optimization and the other related to the surface partitioning strategy optimization. As for the phase shifts optimization, the results derived in Theorem \ref{thm-phase-shifts} can be directly applied, which are not limited to QPSK primary and BPSK secondary constellations.
Then, we substitute the obtained phase shifts into $D_{1}$, $D_{2}$, and $D_{3}$, and the surface partitioning strategy can be obtained by using the following rule
\begin{align} 
    \{N_{1}^{\star},N_{2}^{\star}\}=\arg \max\limits_{N_{1}+N_{2}=N } \{\min\{D_{1},D_{2},D_{3}\}\}. \label{eq:surface-partitioning-extension}
\end{align}

As for~\eqref{eq:surface-partitioning-extension}, we can derive the closed-form solutions similar to our previous procedure, which depend on the modulation constellations of $s$ and $c$.  However, as mentioned in Fig. \ref{fig: without-direct-link}, there may exist several surface partitioning strategies in the candidate set that could maximize the minimum Euclidean distance of the composite signal.
In this case, we can choose best one from the candidate set, which corresponds to a better bit-mapping rule, as discussed in Sec. \ref{sec-Design-criteria}.

In addition, the performance analysis under higher-order modulations can be conducted by following the framework in Sec. \ref{sec-performance-analysis}. Specifically, given the modulation constellations, we can substitute the corresponding Euclidean distances and Hamming distances into \eqref{eq: Px}, \eqref{eq: Ps}, and \eqref{eq: Pc}, and then approximate the BERs with the minimum Euclidean distance derived above, i.e., $\min\{D_1,D_2,D_3\}$. With the approximated BERs, we can analyze the assistance-transmission tradeoff analytically by studying relationship between the minimum Euclidean distance and the surface partitioning strategy.
\vspace{-0.3cm}
\section{Simulation Results} \label{sec-simulation-results}
In this section, simulation results are provided to validate the effectiveness of the proposed RIS partitioning scheme. All channels are composed of large-scale path loss and small-scale fading. We set the simulation parameters with references to the classical RIS papers, e.g., ~\cite{wu2019intelligent}.
Specifically, the path loss is modeled as $\rho=10^{-3}d^{-\xi}$, where $d$ denotes the distance between two nodes and $\xi$ denotes the path loss exponent. The distances of the PTx-CRx link, the PTx-RIS link, and the RIS-CRx link are set to $d_{1}=80 $ $\mathrm{m}$, $d_{2}=75$ $\mathrm{m}$, and $d_{3}=10$ $\mathrm{m}$ with $\xi_{1}=3.5$, $\xi_{2}=2.2$, and $\xi_{3}=2.8$. The small-scale fading of $h_{d}$, $\bm{g}$, and $\bm{h}$ are modeled as Rician fading. Taking $\bm{g}$ as an example, its small-scale fading is given by $\bm{g}_{\rm small}=\sqrt{\frac{\kappa_{g}}{\kappa_{g}+1}}\bm{g}^{\mathrm{LoS}}+\sqrt{\frac{1}{\kappa_{g}+1}}\bm{g}^{\mathrm{NLoS}}$, where $\kappa_{g}$ is the Rician factor; the LoS component $\bm{g}^{\mathrm{LoS}}$ is generated using the steering vectors~\cite{wu2019intelligent}; the non-LoS (NLoS) component $\bm{g}^{\rm NLoS}$ is drawn from the distribution of $\mathcal{CN}(0,1)$. The small-scale fading of $\bm{h}$ and $h_{d}$ can be written in a similar way. We set $\kappa_{g}=10$, $\kappa_{h}=8$, $\kappa_{hd}=12$.
Besides, the noise power is set to $\sigma^{2}=-100$ $\mathrm{dBm}$.
All results are averaged over $10^{6}$ channel realizations.
\vspace{-0.4cm}
\subsection{Assistance-Transmission Tradeoff}
\vspace{-0.1cm}
\begin{figure}[!t]
	\centering 
	\setlength{\abovecaptionskip}{-0.1cm}
	\captionsetup{font={scriptsize}}
	\subfigure[ In the absence of the direct link.] {
		\label{fig:fig_tradeoff_wo_D}
		\includegraphics[width=0.4\textwidth]{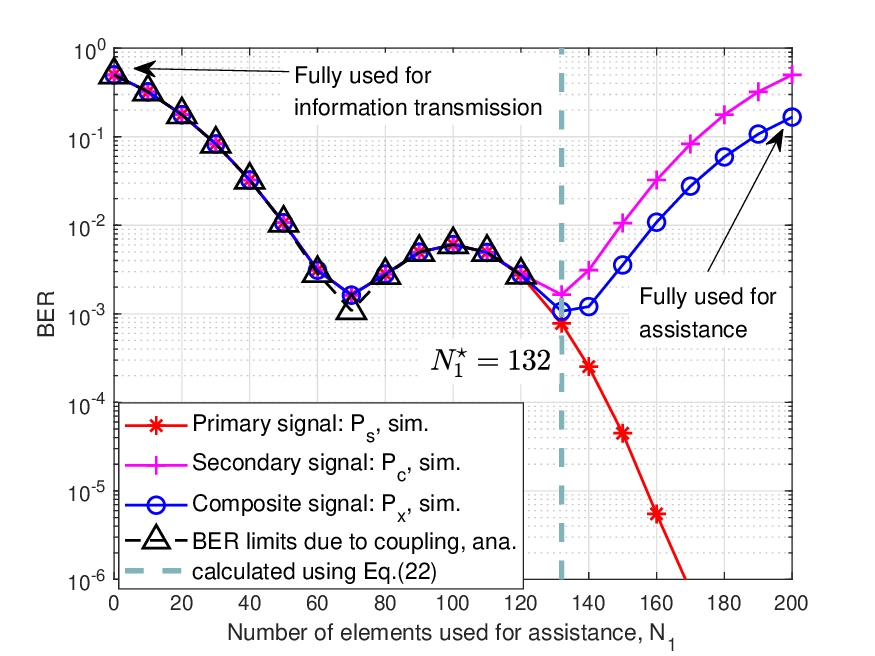}
	}
	\subfigure[In the presence of the direct link.]{
		\label{fig:fig_tradeoff_w_D}
		\includegraphics[width=0.4\textwidth]{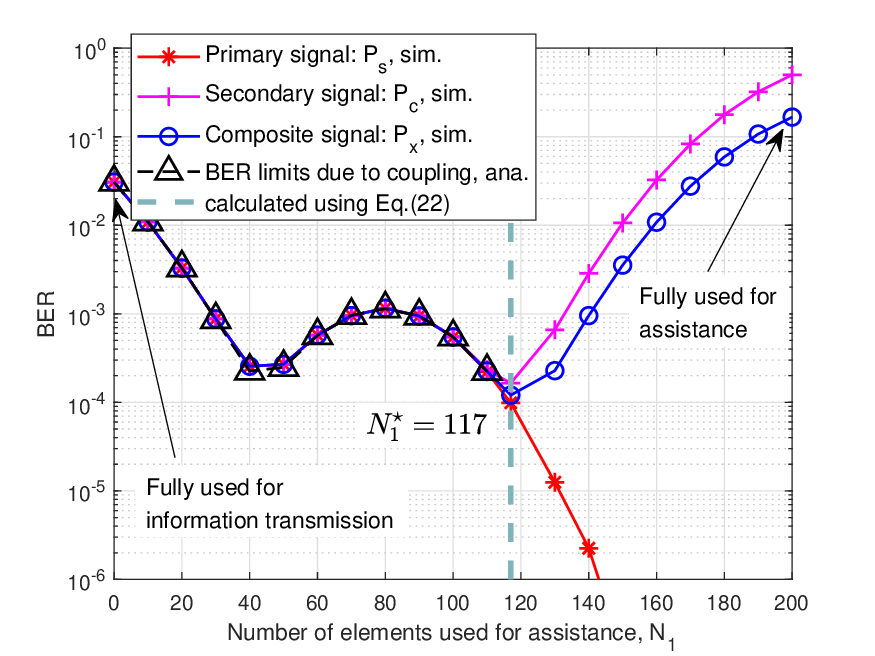}
	}
	\caption{ $P_{x}$, $P_{s}$, and $P_{c}$ vs. number of elements used for assistance, with total number of elements $N=200$.}\label{fig: fig_tradeoff}
\end{figure}

To verify the effectiveness of our proposed tradeoff methodology, we plot the BERs of signals $s$, $c$, and $x$ versus the number of elements used for assistance in Fig. \ref{fig: fig_tradeoff}. First, we consider the scenario where the direct link is blocked and the BERs are shown in Fig. \ref{fig: fig_tradeoff} (a). In this scenario, the optimal $N_{1}^{\star}=132$ is calculated using \eqref{eq: optimal_partitioning_strategy} with $t=0$. It is observed that under the optimized partitioning strategy, the BER of the composite signal, $P_{x}$, is minimized. This indicates that the analysis in Theorem \ref{thm-elements-allocation} under the LoS channels also shows a good performance for Rician fading. Besides, we also observe that for $0<N_{1}<120$, the coupling effect dominates their BER and thus there exists the BER limits $\mathcal{Q}(\sqrt{\frac{p_{t}D_{1}^{2}}{2\sigma^{2}}})$ and $\mathcal{Q}(\sqrt{\frac{p_{t}D_{2}^{2}}{2\sigma^{2}}})$, where $D_{1}=2\sum_{n=1}^{N_{1}}|h_{1,n}g_{1,n}|$, $D_{2}=\sqrt{2}|\sum_{n=1}^{N_{1}}|h_{1,n}g_{1,n}|+\sum_{n=1}^{N_{2}}|h_{2,n}g_{2,n}e^{\jmath \frac{3}{4}\pi}||$ are the minimum Euclidean distances when $0\leq N_{1}\leq 70$ and $70\leq N_{1} \leq 120$, respectively. This validates the accuracy of our coupling analysis in Sec. \ref{sec-performance-analysis}.
When $N_{1}$ exceeds $N_{1}^{\star}$, $P_{s}$ monotonically decreases while $P_{c} $ increases. Thus, the partitioning strategy under $N_{1}^{\star}$ achieves the best assistance-transmission tradeoff.

\begin{figure}[t]
	\centering
	\captionsetup{font={scriptsize}}
	\setlength{\abovecaptionskip}{-0.07cm}
	\begin{minipage}[t]{0.4\textwidth}
		\centering  
		\includegraphics[width=2.8in]{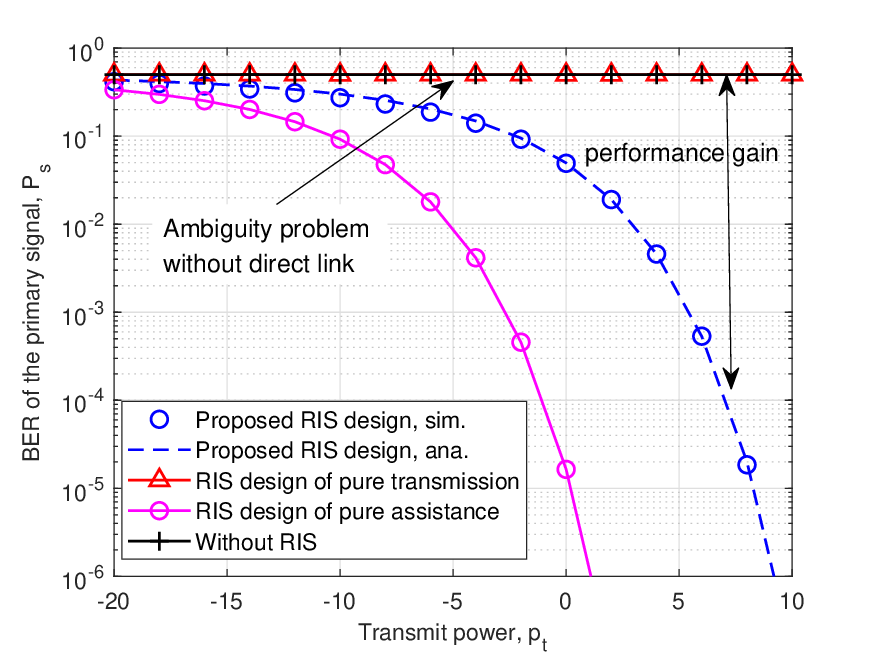} 
		\caption{BER of primary signal $P_{s}$ vs. power $p_{t}$ w/o direct link }  
		\label{fig:fig_Ps_wo_D}   
	\end{minipage}
	\begin{minipage}[t]{0.4\textwidth}
		\centering  
		\includegraphics[width=2.8in]{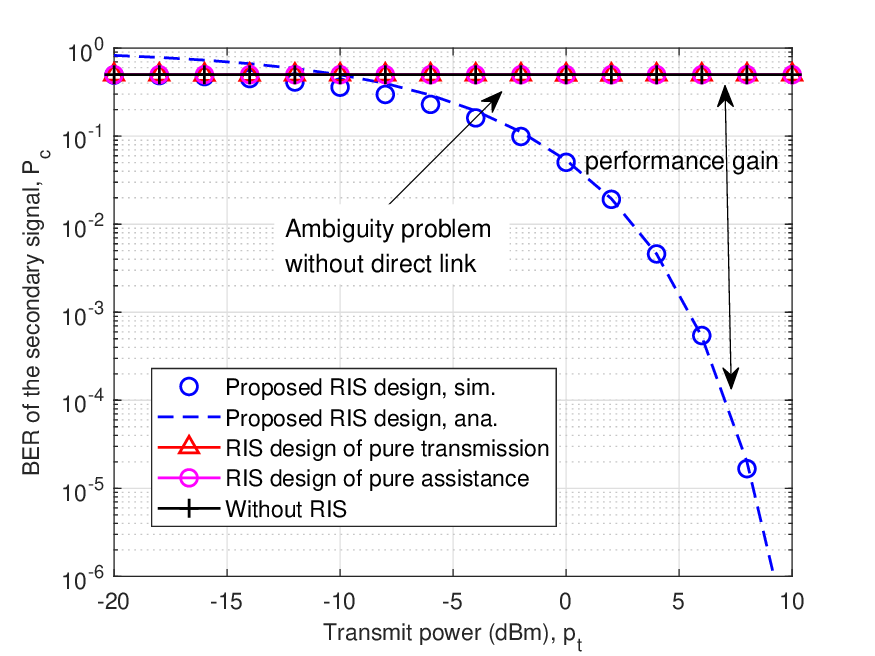} 
		\caption{BER of secondary signal $P_{c}$ vs. power $p_{t}$ w/o direct link}
		\label{fig:fig_Pc_wo_D}   
	\end{minipage}
\end{figure}
\begin{figure}[t]
	\centering
	\captionsetup{font={scriptsize}}
	\setlength{\abovecaptionskip}{-0.07cm}
	\begin{minipage}[t]{0.4\textwidth}
		\centering  
		\includegraphics[width=2.8in]{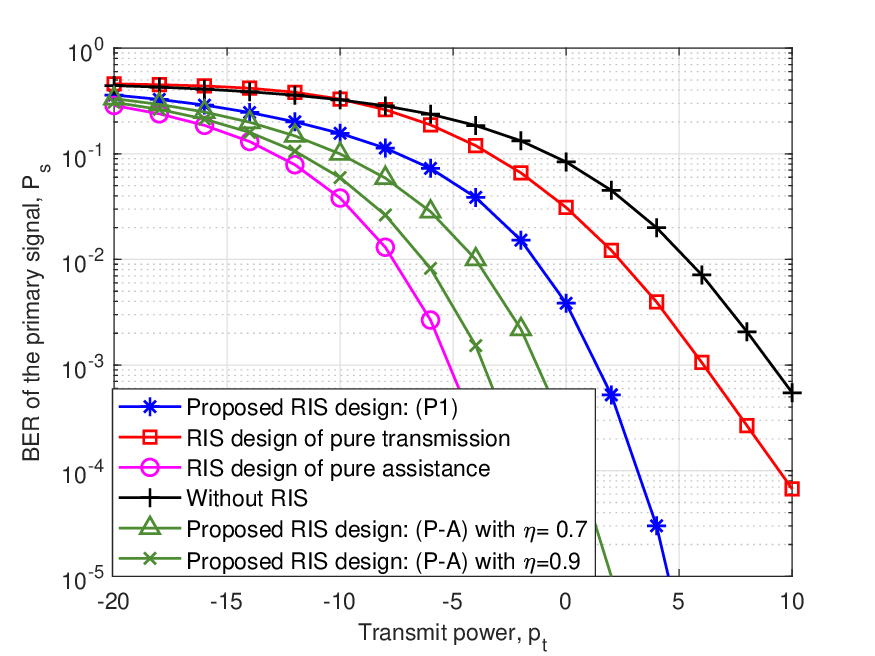} 
		\caption{BER of primary signal $P_{s}$ vs.  power $p_{t}$ w/ direct link }  
		\label{fig:fig_Ps_w_D}   
	\end{minipage}
	\begin{minipage}[t]{0.4\textwidth}
		\centering  
		\includegraphics[width=2.8in]{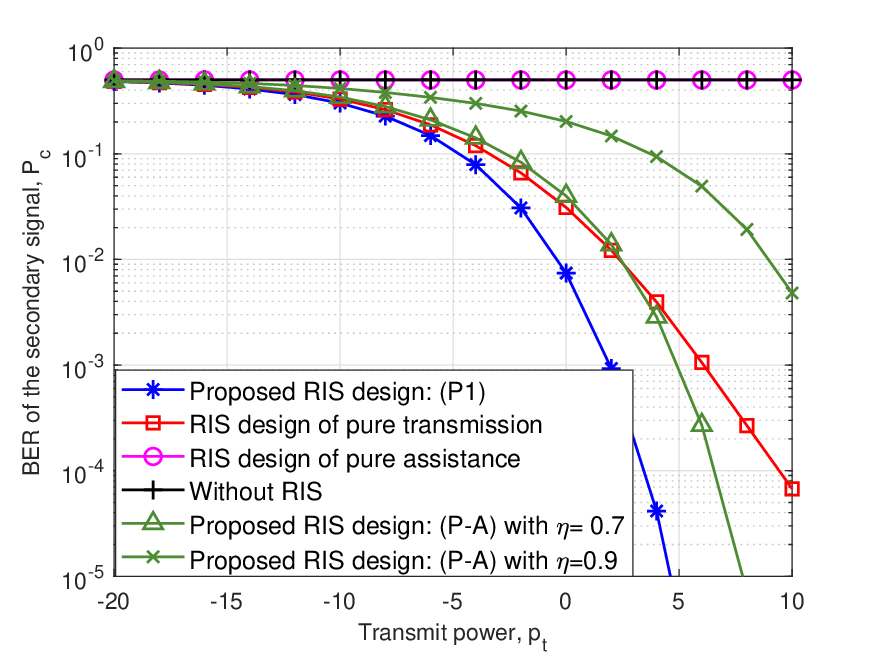} 
		\caption{BER of secondary signal $P_{c}$ vs. power $p_{t}$ w/ direct link}
		\label{fig:fig_Pc_w_D}   
	\end{minipage}
\end{figure}
Furthermore, in Fig. \ref{fig: fig_tradeoff} (b), we plot the BERs under the general scenario where the direct link exists. The optimal $N_{1}^{\star}=117$ is calculated using \eqref{eq: optimal_partitioning_strategy} with $t=\sqrt{\frac{\rho_{1}}{\rho_{2}\rho_{3}N^2}}=0.21$ under LoS channels.
Notably, when the direct link exists, the optimized number of elements allocated for assistance is less than the scenario without the direct link, i.e., $117<132$. This is because the existence of the direct link can innately enhance the primary transmission, and thus it requires fewer RIS elements for assistance. 
Moreover, in Fig. \ref{fig: fig_tradeoff} (b), we validate the commutative property when the direct link exists. Through calculation, the strength of the direct link approximately equals the strength of the reflected link when RIS is equipped with $N_{d}=40$ elements since $\sqrt{\frac{\rho_{1}}{\rho_{2}\rho_{3}}}\approx 40$. By leveraging Remark \ref{rema: commutative}, it is equivalent to say that RIS has $N+N_{d}=200+40=240$ elements in total and the RIS allocates $40$ elements for assistance in advance. 
With this knowledge, taking $N_{1}=60$ and $N_{1}=100$ in Fig. \ref{fig: fig_tradeoff} (b) as an example, it is found that we can either use the scheme (assistance: $(60+40)$ elements, transmission: $140$ elements) or the other scheme (assistance: $(100+40)$ elements, transmission: $100$ elements) to achieve the similar BER performance of the primary and secondary signals. Thus, the commutative property still holds when the direct link exists.
\vspace{-0.4cm}
\subsection{Validity of Proposed RIS Partitioning Scheme}
To reveal more insights into the assistance-transmission tradeoff, we consider the following benchmarks
\begin{itemize}
    \item \textbf{Benchmark 1 (Without RIS)}: The RIS-assisted SR system degrades to conventional point-to-point communication between the PTx and the CRx.
    \item \textbf{Benchmark 2 (RIS design of pure assistance)}\cite{wu2019intelligent}: All the RIS elements are used to assist the primary transmission, where $N_{1}=N$, $N_{2}=0$.
    \item \textbf{Benchmark 3 (RIS design of pure transmission}\cite{zhou2022cooperative,hua2021uav,zhang2021reconfigurable,hua2021novel,hu2020reconfigurable}: All the RIS elements are used to transmit the secondary signal, where $N_{1}=0$, $N_{2}=N$.
\end{itemize}

Considering the case of a blocked direct link, in Fig. \ref{fig:fig_Ps_wo_D} and Fig. \ref{fig:fig_Pc_wo_D}, we plot the BER performance versus the transmit power under different benchmarks. Note that for Benchmark 1 and Benchmark 2, RIS can not transmit the secondary signal, and here we let its BER equal $0.5$, i.e., $P_{c}=0.5$. One special phenomenon is the ambiguity problem under the conventional RIS design of pure transmission, which results in $P_{s}=P_{c}=0.5$ regardless of transmit power. However, by adopting our proposed RIS design, the ambiguity problem can be solved since part of RIS elements is allocated for assistance, which can be viewed as a virtual direct link. Moreover, we see that the analytical BER under the proposed scheme is consistent with the simulations and the performance gain increases with the transmit power, which corroborates the accuracy of the performance analysis in Sec. \ref{sec-performance-analysis}. 

Notably, from Fig. \ref{fig:fig_Ps_wo_D}, there exists a performance gap of the primary transmission under the proposed design and Benchmark 2. 
To reduce such a gap, \textbf{P-A} is one way to guarantee the performance requirement of the primary transmission. In Fig. \ref{fig:fig_Ps_w_D}, we plot the BERs of the primary and secondary signals under different requirements $D_{s,0}$ in \textbf{P-A}, when the direct link exists.
Here, we introduce a factor $ \eta \in[0, 1]$ and set the requirement as $D_{s,0}=\eta D_{s,\max}$, where $D_{s,\max}$ denotes the minimum Euclidean distance when RIS is designed to purely assist the primary transmission. From Fig. \ref{fig:fig_Ps_w_D}, we see that with the increase of $\eta$ in \textbf{P-A}, the primary transmission can achieve a higher BER performance than the proposed RIS design by solving \textbf{P1}. Therefore, the priority of the primary transmission can be guaranteed by controlling the factor $\eta$, where $\eta=1$ means RIS needs to purely assist the primary transmission, i.e., Benchmark 2.

On the other hand, from Fig. \ref{fig:fig_Pc_w_D}, it is observed that when we increase $\eta$ in \textbf{P-A}, the performance of the secondary transmission will be degraded as compared to our proposed design by solving \textbf{P1}. This is due to the fact that increasing $\eta$ will force RIS to allocate more reflecting elements for assistance. However, in this case, less reflecting elements are left for transmission, which leads to a worse performance of the secondary transmission.

\vspace{-0.4cm}
\subsection{Impact of Total Number of RIS Reflecting Elements}
\begin{figure}[t]
	\centering
	\captionsetup{font={scriptsize}}
	\begin{minipage}[t]{0.48\textwidth}
		\centering  
		\includegraphics[width=2.8in]{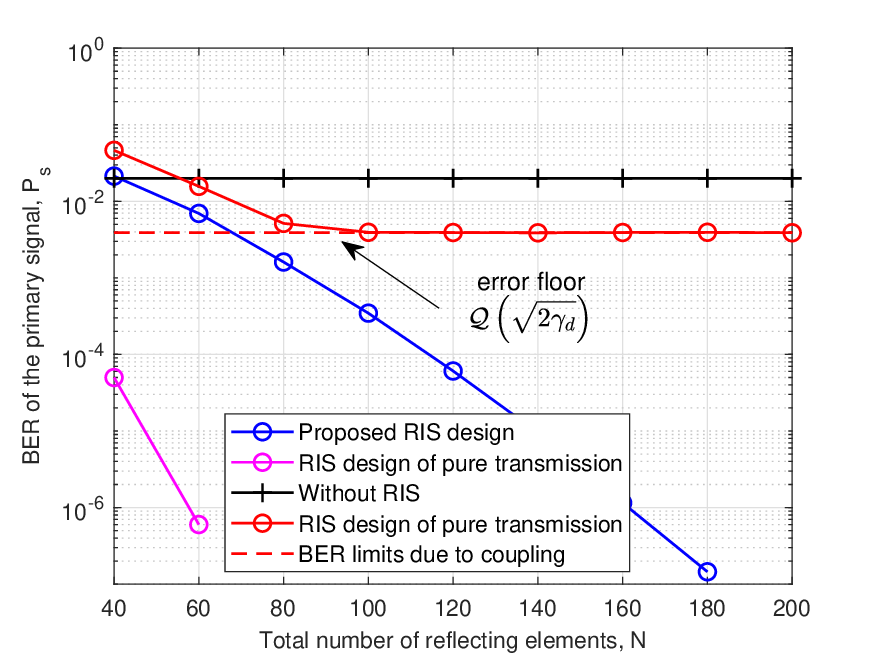} 
		\caption{BER of primary signal $P_{s}$ vs.  total number of elements }  
		\label{fig:fig_Ps_w_D_number}   
	\end{minipage}
	\begin{minipage}[t]{0.48\textwidth}
		\centering  
		\includegraphics[width=2.8in]{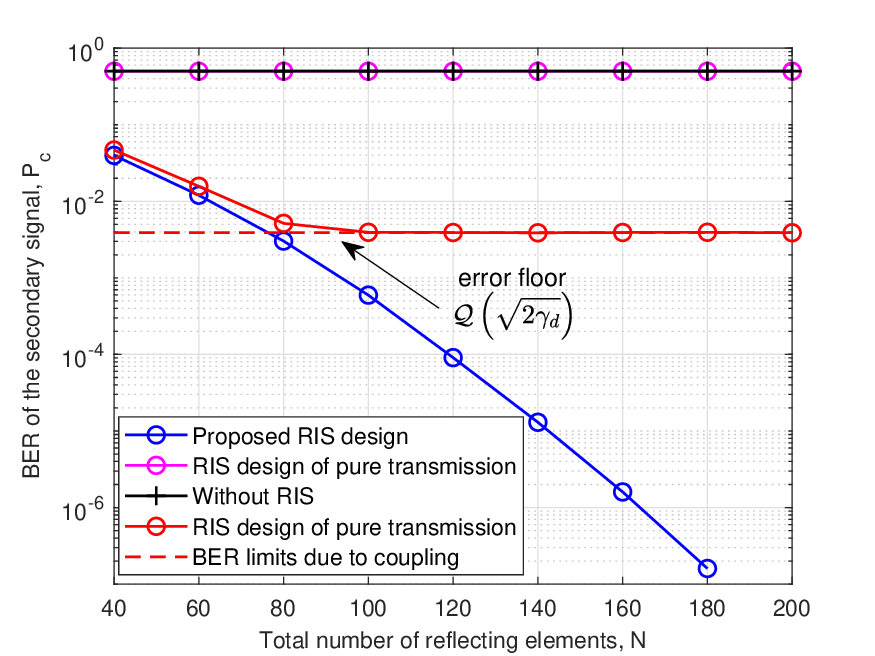} 
		\caption{BER of secondary signal $P_{c}$ vs. total number of elements}
		\label{fig:fig_Pc_w_D_number}   
	\end{minipage}
\end{figure}
\begin{figure}[t]  
	\centering  
	\captionsetup{font={scriptsize}}
	\includegraphics[width=2.8in]{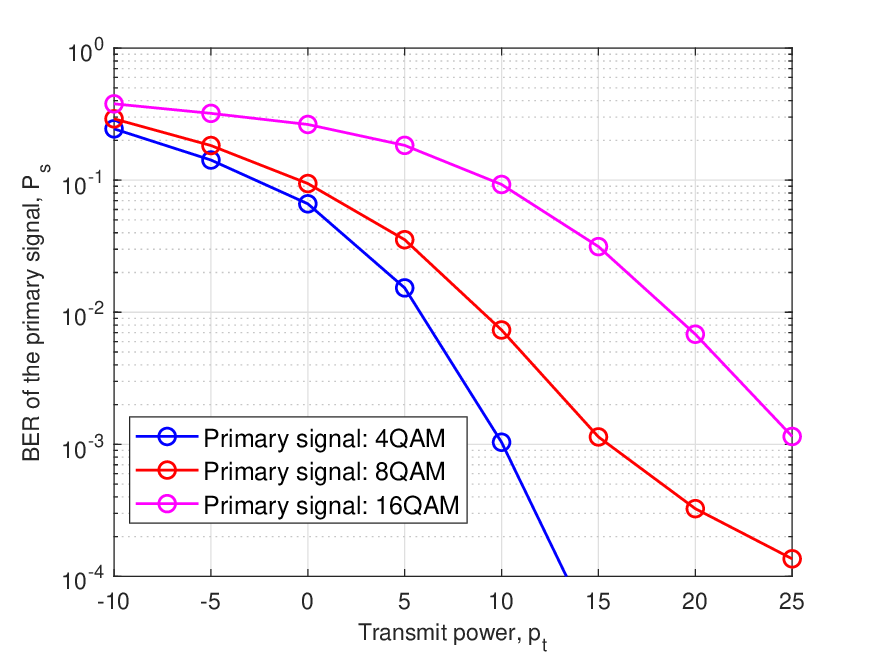}  
	\caption{BER of primary signal vs. transmit power under different $M_s$-ary QAM modulation, for a fixed QPSK secondary signal.}
 \vspace{-0.4cm}
	\label{fig:primary-fix-secondary}  
\end{figure} 
\begin{figure}[t]  
	\centering  
	\captionsetup{font={scriptsize}}
	\includegraphics[width=2.8in]{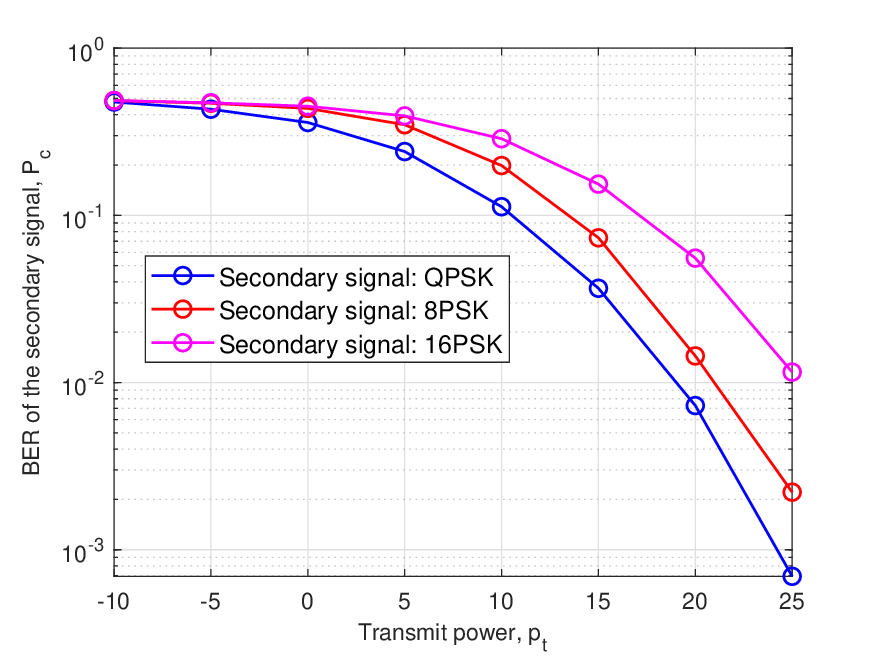}  
	\caption{BER of secondary signal vs. the transmit power under different $M_c$-ary PSK modulation, for a fixed 16QAM primary signal.}
	\label{fig:secondary-fix-primary}  
\end{figure}
Next, in Fig. \ref{fig:fig_Ps_w_D_number} and Fig. \ref{fig:fig_Pc_w_D_number}, we plot the BERs of the primary and secondary signals versus the total number of RIS elements. It is observed that with the increase in $N$, conventional RIS design, i.e., Benchmark 3, suffers from the error-floor problem, i.e., the BERs of the primary and secondary signals converge to a non-zero value, $\mathcal{Q}(\sqrt{2\gamma_{d}})$, irrespective of $N$, where $\gamma_{d}=\frac{p_{t}|h_{d}|^2}{\sigma^2}$ is defined as the SNR of the direct link. This phenomenon can be explained as follows. When $N$ goes to infinity, the reflected link will be very strong compared to the direct link, and then the coupling caused by signal multiplication will dominate their BERs as shown in Sec. \ref{sec: coupling effect}. By using our proposed RIS design scheme, the error-floor problem can be effectively addressed. This is because our proposed scheme assigns part of RIS elements to assist the primary transmission, which strengthens the equivalent direct link and reduces the impact of the coupling effect, thereby achieving an error-floor free state. Interestingly, from Fig. \ref{fig:fig_Ps_w_D_number}, it can be seen that when $N$ is larger than $40$, the BER of the primary transmission becomes better than the case without RIS.
This motivates us to figure out the minimum number of reflecting elements, under which the primary transmission can gain benefits from RIS. Specifically, we can compare the Euclidean distance of the primary signal with the case where RIS is absent. Take the case 1 in Theorem \ref{thm-elements-allocation} as an example. In this case, the minimum Euclidean distance of the primary signal is $2|h_{r,2}|$ as analyzed in Fig. \ref{fig: fig_tradeoff} (a), while the minimum Euclidean distance of the primary signal without RIS is $\sqrt{2}|h_d|$. By solving the equality $2|h_{r,2}|=\sqrt{2}|h_d|$, where $h_{r,2}=\sqrt{\rho_{2}\rho_{3}}$ and $h_d=\sqrt{\rho_{1}}$ under LoS channels, we can immediately obtain the solution $N=\left \lceil\frac{\sqrt{3}-1}{\sqrt{2}+1-\sqrt{3}}\sqrt{\frac{\rho_{1}}{\rho_{2}\rho_{3}}}\right \rceil$. It is shown that if the direct link is stronger, more reflecting elements are required to ensure that the primary transmission can be enhanced due to the existence of the RIS.

\vspace{-0.3cm}
\subsection{Performance of Higher-Order Modulation Constellations}
In this subsection, we extend our design and analysis to the cases of higher-order modulation constellations. Specifically, we consider $M_{s}$-ary QAM and $M_{c}$-ary PSK constellations for primary and secondary signals, respectively. In Fig. \ref{fig:primary-fix-secondary}, we plot the BER of the primary signal under different $M_s = 4,8,16$, for a fixed QPSK secondary signal. It can be observed that with the increase of $M_{s}$, primary BER performance will be degraded. For example, when $P_{s}=10^{-3}$, 8QAM constellations requires more than $5$ dB tranmit power to achieve the same performance as the 4QAM constellation.
This is obvious since a higher-order QAM signal has a smaller Euclidean distance under the constraint of unit average power.
Similarly, in Fig. \ref{fig:secondary-fix-primary}, we plot the BER of the secondary signal under different $M_{c}=4,8,16$, for a fixed $16$QAM primary signal. It is also shown that with the increase of PSK modulation order, the BER performance of the secondary transmission will be degraded. When $P_{c}=10^{-3}$, 16PSK constellation requires more than $7$ dB transmit power to achieve the same performance as the QPSK constellation. 

\begin{figure}[t]  
	\centering  
	\captionsetup{font={scriptsize}}
	\includegraphics[width=2.8in]{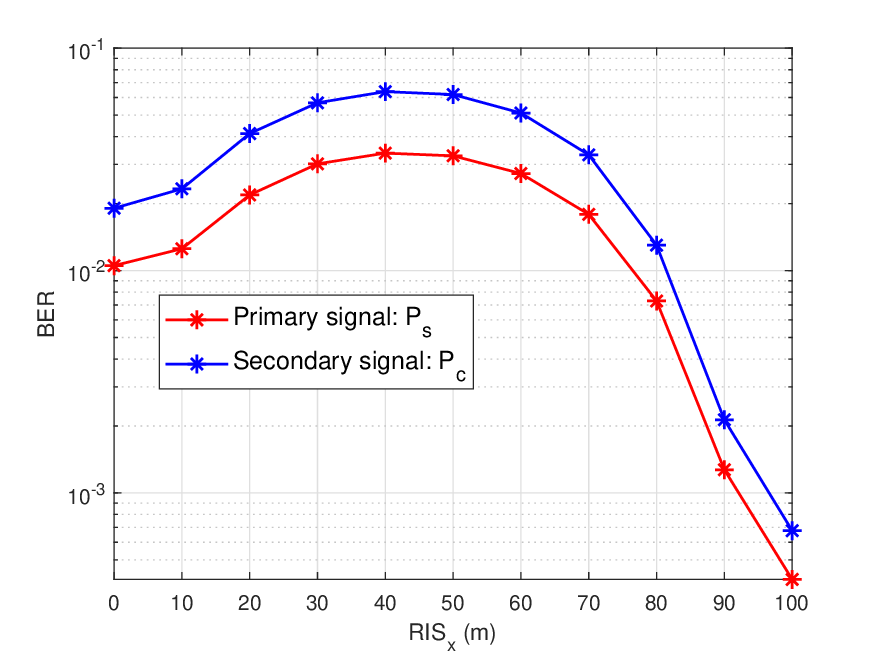}  
	\caption{BER of primary and secondary signals vs. the location of RIS.}
	\label{fig:BER-distance}  
 \vspace{-0.4cm}
\end{figure} 
\begin{figure}[t]  
	\centering  
	\captionsetup{font={scriptsize}}
	\includegraphics[width=2.8in]{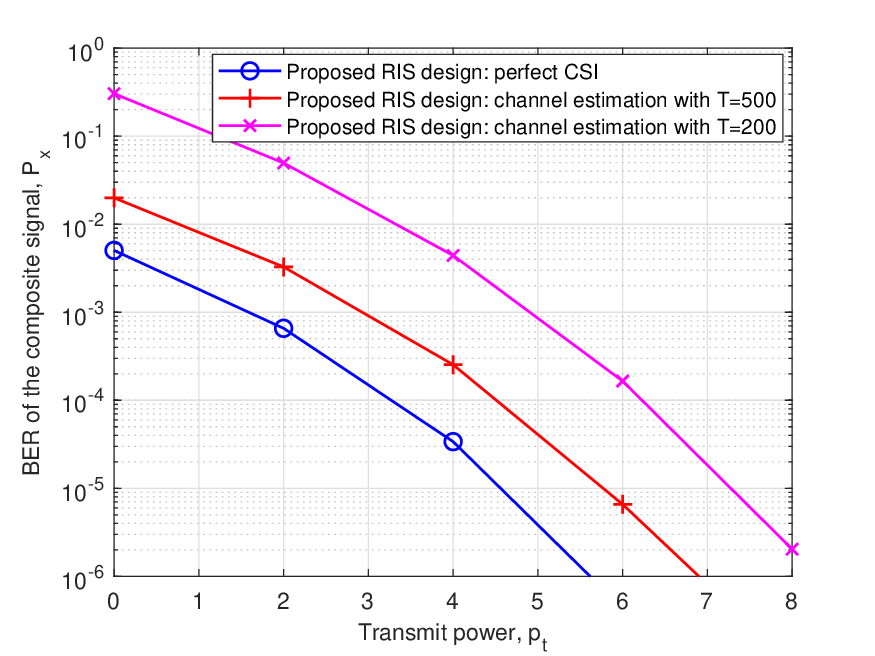}  
	\caption{BER of primary and secondary signals vs. transmit power under different training time.}
	\label{fig:BER-framestructure}  
\end{figure} 

\vspace{-0.3cm}
\subsection{Impact of Node Distances and Frame Structure Design on System Performance}\label{sec-frame}
In this subsection, we study the effects of distances among PTx, RIS, and CRx on the system performance. Following~\cite{wu2019intelligent}, we consider a two-dimensional setup, where the coordinates of the PTx and the CRx are set to $(0,0)$ and $(100 ,0)$, respectively. Besides, the coordinate of the RIS is set to $(\mathrm{RIS}_{x},15)$, where $\mathrm{RIS}_{x}$ varies from $0$ to $100 (\mathrm{m})$. Then, we plot the BER of the primary and secondary signals when $\mathrm{RIS}_{x}\in [0,100]$ in Fig. \ref{fig:BER-distance}. It is shown that $P_{s}$ and $P_{c}$ achieve their worst performance when RIS is located at the middle of the PTx and the CRx, while BER performance becomes better when RIS is deployed closer to the PTx or the CRx. This is due to the fact that the product path-loss of the reflected link becomes larger when RIS is placed at the middle.
Moreover, under our setting, it is better to deploy RIS closer to the CRx. This is because the path loss exponent of the RIS-CRx link is larger than that of the PTx-RIS link, which implies a smaller path loss of the cascaded PTx-RIS-CRx link when RIS is close to the CRx.  

\begin{table}[t]
\centering 
\caption{The Main Parameters of the Frame Strucure}
\begin{tabular}{|c|c|}
\hline
\textbf{Parameter}     & \textbf{Value} \\ \hline
Modulation             & QPSK           \\ \hline
Transmission rate, $R_b$      & 10 Mbps        \\ \hline
Symbol period, $T_b$          & 0.2 us         \\ \hline
Channel coherence time, $T_a$ & 2.5 ms         \\ \hline
\end{tabular} \label{table1}
\end{table}

Next, we study the impact of frame structure design on the system performance. The main parameters of the frame structure are summarized in Table \ref{table1}. We consider that channel coherence time is $T_a=2.5 (\mathrm{ms})$ with reference to~\cite{tse2005fundamentals}, corresponding to a slow-fading channel scenario.
Assume primary transmission rate is $10 (\mathrm{Mbps})$ with QPSK modulation. Then its symbol period can be calculated as $T_{b}=0.2 (\mathrm{us})$. As illustrated in Fig. \ref{fig:system-model} (b), the frame structure consists of channel estimation and data transmission. 
The training length required for channel estimation is denoted by $T$. Here, we take the least square estimation~\cite{swindlehurst2022channel} as an example to show the impact of training length on the overall BER performance of RIS-assisted SR, which is reflected by the BER of the composite signal.
In Fig. \ref{fig:BER-framestructure}, we plot the BER of the composite signal versus the transmit power under different training length. It can be seen that with the increasing in training time $T$, the BER performance becomes better and gradually approaches the case with the perfect CSI due to the fact that a longer training leads to a more accurate channel estimation at the cost of shorter time for data transmission.
\vspace{-0.4cm}
\section{Conclusion}
\label{sec-conslusion}
In this paper, we have proposed a novel RIS partitioning scheme for SR, which provides more degrees of freedom to balance the assistance and transmission capabilities of RIS. To achieve the best assistance-transmission tradeoff, we have formulated the problem to minimize the BER of the composite signal by jointly optimizing the surface partitioning strategy and the phase shifts.
By solving this problem, we have shown that there indeed exists one partitioning strategy, under which the BER of the composite signal is minimized and the best tradeoff is thus achieved. After that, performance analysis has been conducted to draw insights into RIS-assisted SR, including the coupling effect analysis, the performance gain, and the commutative property of assistance-transmission dual capabilities. Finally, simulation results have shown that our proposed scheme outperforms the conventional schemes which design RIS for either assistance or transmission.
\vspace{-0.4cm}
\begin{appendices} 
\section{}\label{appendix-phase-shifts}
{ Let $\bm{\phi}_{1}=[\phi_{1,1},\cdots,\phi_{1,N_{1}}]^{H}$, $\bm{\phi}_{2}=[\phi_{2,1},\cdots,\phi_{2,N_{2}}]^{H}$, $\bm{f}_{1}=\mathrm{diag}(\bm{h}_{1}^{H})\bm{g}_{1}$, $\bm{f}_{2}=\mathrm{diag}(\bm{h}_{2}^{H})\bm{g}_{2}$. 
According to the relationship between $\{s_{m},c_{m}\}$ and $\{s_{k},c_{k}\}$, the squared Euclidean distance can be classified into three cases.
\begin{itemize}
        \item \emph{Case 1}: $s_{m}\neq s_{k}$, $c_{m}= c_{k}$. For this case, we have
    \begin{align}
        D_{m,k}^{2}
        &=  (|h_{d}+\bm{\phi}_{1}^{H}\bm{f}_{1}|^2+|\bm{\phi}_{2}^{H}\bm{f}_{2}|^2 
         \nonumber \\
        &
        +2\Re\{(h_{d}+\bm{\phi}_{1}^{H}\bm{f}_{1})^{H}\bm{\phi}_{2}^{H}\bm{f}_{2}c_{m}\}) |s_{m}-s_{k}|^2 . \nonumber
    \end{align}
    \item \emph{Case 2}: $s_{m}=s_{k}$, $c_{m}\neq c_{k}$. In this case, we have
    \begin{align} \nonumber
        D_{m,k}^{2}= \left|\bm{\phi}_{2}^{H}\bm{f}_{2}\right|^2|c_{m}-c_{k}|^2.
    \end{align}
    \item \emph{Case 3}: $s_{m}\neq s_{k}$, $c_{m}\neq c_{k}$. For this case, we have
    \begin{align}
        D_{m,k}^{2}&= |h_{d}+\bm{\phi}_{1}^{H}\bm{f}_{1}|^2|s_{m}-s_{k}|^2
       +
        |\bm{\phi}_{2}^{H}\bm{f}_{2}|^2 \nonumber \\ &\quad |s_{m}c_{m}-s_{k}c_{k}|^2  +     2\Re\{(h_{d}+\bm{\phi}_{1}^{H}\bm{f}_{1})^{H}\bm{\phi}_{2}^{H}\bm{f}_{2}\} \nonumber \\ &\quad  (s_{m}-s_{k})^{H}(s_{m}c_{m}-s_{k}c_{k}). \nonumber
    \end{align}
\end{itemize}

From the above three cases, we have three observations regarding maximizing the minimum Euclidean distance, which are listed as follows.

\emph{i):} The phase shifts of the sub-surface \uppercase\expandafter{\romannumeral1}, $\bm{\phi}_{1}$, should be designed to align the reflected link via sub-surface \uppercase\expandafter{\romannumeral1}, i.e., $\bm{\phi}_{1}^{H}\bm{f}_{1}$,  with the direct link $h_{d}$, such that the Euclidean distance term $|h_{d}+\bm{\phi}_{1}^{H}\bm{f}_{1}|^2$ under \emph{Case 1} and \emph{Case 3} is maximized.

\emph{ii):} The phase shifts of the sub-surface \uppercase\expandafter{\romannumeral2}, $\bm{\phi}_{2}$, should be designed to align the reflected link via sub-surface \uppercase\expandafter{\romannumeral2}, i.e., $\bm{\phi}_{2}^{H}\bm{f}_{2}$, such that the Euclidean distance under \emph{Case 2} is maximized.

\emph{iii):} Based on the results obtained from \emph{i)} and \emph{ii)}, additional common phase shift $\overline{\phi}$ is required for $\bm{\phi}_{2}$ to maximize the term $2\Re\{(h_{d}+\bm{\phi}_{1}^{H}\bm{f}_{1})^{H}\bm{\phi}_{2}^{H}\bm{f}_{2}\}(s_{m}-s_{k})^{H}(s_{m}c_{m}-s_{k}c_{k})$ for all possible pairs of $\{s_{m},c_{m}\}$ and $\{s_{k},c_{k}\}$, such that the minimum Euclidean distance under \emph{Case 1} and \emph{Case 3} is maximized. Note that the Euclidean distance under \emph{Case 2} can still be maximized even though the additional common phase shift $\overline{\phi}$ is added to the sub-surface \uppercase\expandafter{\romannumeral2}.

Combining all the three observations, Theorem \ref{thm-phase-shifts} follows.}
\vspace{-0.4cm}
\section{}\label{appendix-partitioning-condition}
\begin{figure}[t]  
	\centering  
	\captionsetup{font={scriptsize}}
	\setlength{\abovecaptionskip}{-0cm}
	\includegraphics[width=3.2in]{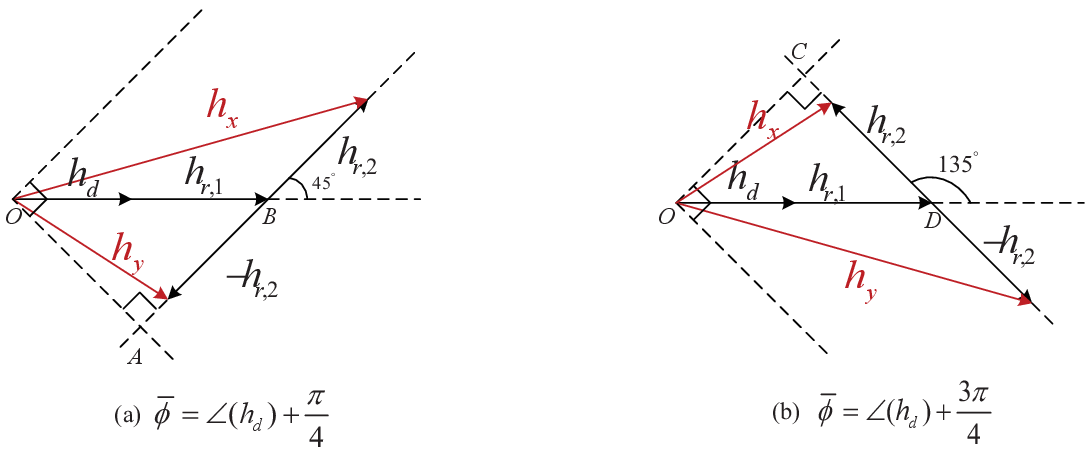}  
	\caption{Geometrical illustration of Lemma \ref{lem-constraint}.} 
	\label{fig:phase-rotation}  
\end{figure}
Let $h_{r,1}=(\bm{\phi}_{1}^{\star})^{H}\bm{f}_{1}=\sum_{n=1}^{N_{1}}|h_{1,n}||g_{1,n}|e^{\jmath \angle(h_{d})}$, $h_{r,2}=(\bm{\phi}_{2}^{\star})^{H}\bm{f}_{2}=\sum_{n=1}^{N_{2}}|h_{2,n}||g_{2,n}|e^{\jmath\overline{\phi}}$, where $\overline{\phi}=\angle(h_{d})+\frac{\pi}{4}$ or $\angle(h_{d})+\frac{3\pi}{4}$. Denote the composite channels by $h_{x}=h_{d}+h_{r,1}+h_{r,2}$ and $h_{y}=h_{d}+h_{r,1}-h_{r,2}$, when $c=1$ and $c=-1$, respectively. 
As illustrated in Fig. \ref{fig:phase-rotation}, we sketch these composite channels when $\overline{\phi}$ take different values. Take Fig. \ref{fig:phase-rotation} (a) as an example. It is observed no matter how we vary $N_{1}$ and $N_{2}$, the phase of the composite channel $h_{x}$, i.e., $\angle(h_{x})$, always satisfies the constraint $\mathrm{C1}$. Then, we focus on the transition point when $\mathrm{C2}$ is active.
It is observed that if we increase $N_{2}$, when $h_{y}$ equals $\overrightarrow{OA}$, the phase of $h_{y}$ is then $\angle(h_{y})=\angle(h_{d}-\frac{\pi}{4})$. In this case, we have $|h_{d}|+|h_{r,1}|=\sqrt{2}|h_{r,2}|$. 
For the case in Fig. \ref{fig:phase-rotation} (b), the same results can also be obtained.
\vspace{-0.5cm}
\section{}
\label{appendix-elements-allocation}
\vspace{-0.2cm}
By dropping the irrelevant terms, the squared Euclidean distances are given by  $f_{1}(N_{1})=2\rho_{2}\rho_{3}\times [(2+\sqrt{2})N_{1}^{2}-(2+\sqrt{2})(1-t)N N_{1}+N^{2}(1-\sqrt{2}t+t^2)]$, where $t=\sqrt{\frac{\rho_{1}}{\rho_{2}\rho_{3}N^2}}$, $f_{2}(N_{1})=(2(\sqrt{\rho_{1}}+\sqrt{\rho_{2}\rho_{3}}N_{1}))^2$, $f_{3}(N_{1})=(2\sqrt{\rho_{2}\rho_{3}}(N-N_{1}))^2$. Next, by discussing the monotonicity of these squared distances, we find that the squared form of the minimum Euclidean distance, i.e., $D_{\min}^{LoS}(N_{1})=\min\{f_{1}(N_{1}),f_{2}(N_{1}), f_{3}(N_{1})\}$, increases with $N_{1}$ when $N_{1}\in [0, \widetilde{N}_{12}]$ and $N_{1}\in [\widetilde{N}_{0},\widetilde{N}_{13}]$, and decreases with $N_{1}$ when $N_{1} \in [\widetilde{N}_{12},\widetilde{N}_{0}]$ and $N_{1}\in [\widetilde{N}_{13},N]$, where $\widetilde{N}_{0}=\frac{1-t}{2}N$, $\widetilde{N}_{12}=\frac{\sqrt{2}+1-\sqrt{3}+(\sqrt{2}-1-\sqrt{3})t}{2}N$ and $\widetilde{N}_{13}=\frac{(\sqrt{3}+1-\sqrt{2})+(\sqrt{3}-1-\sqrt{2})t}{2}N$.
Next, by considering the zeros of $\widetilde{N}_{0}=0$, $\widetilde{N}_{1}=0$, and $\widetilde{N}_{13}=0$, i.e., $t=1$, $t=\sqrt{2}$, and $t=\frac{\sqrt{2}+\sqrt{6}}{2}$, the parameter $t$ is partitioned into four intervals. Then, we will discuss the optimal $N_{1}^{\star}$ in these intervals.

 \emph{Case 1}: $0\leq t\leq 1$. In this case, the axis of symmetry of $f_{1}(N_{1})$, $\widetilde{N}_{0}$, is larger than zero and $\widetilde{N}_{1}>\widetilde{N}_{0}$. The feasible region of $N_{1}$ is the interval $[\widetilde{N}_{1},N]$.
    By solving the equality $\widetilde{N}_{1}=\widetilde{N}_{13}$, we obtain the solution $\widetilde{t}_{1}=\sqrt{3}-\sqrt{2}$ and $\widetilde{t}_{2}=\sqrt{3}(\sqrt{2}-1)$. Then, it follows that
    \begin{itemize}
        \item \emph{Subcase 1}: $0\leq t \leq \sqrt{3}-\sqrt{2}$. We know that $\widetilde{N}_{1} \leq \widetilde{N}_{13}$. The optimal $N_{1}^{\star}$ that maximizes the objective function is $N_{1}^{\star}= \widetilde{N}_{13}=\frac{(\sqrt{3}+1-\sqrt{2})+(\sqrt{3}-1-\sqrt{2})t}{2}N$.
        \item \emph{Subcase 2}: $\sqrt{3}-\sqrt{2} \leq t \leq \sqrt{3}(\sqrt{2}-1) $. We have $\widetilde{N}_{1} \geq \widetilde{N}_{13}$. Thus, $D_{\min}^{LoS}$ monotonically decreases when $N_{1} \in [\widetilde{N}_{1},N]$. We conclude the optimal $N_{1}^{\star}$ is $N_{1}^{\star}=\widetilde{N}_{1}=(2-\sqrt{t^2+2})N$.
        \item \emph{Subcase 3}: $ \sqrt{3}(\sqrt{2}-1) \leq t \leq 1 $. We have $\widetilde{N}_{1}\leq \widetilde{N}_{13}$. Hence, the optimal $N_{1}^{\star}$ is also $N_{1}^{\star}=\widetilde{N}_{13}$.
    \end{itemize}
The other cases can be proved in a similar way. By combining all the cases, Theorem \ref{thm-elements-allocation} follows.
\vspace{-0.4cm}
\section{} \label{appendix-performance-gain}
To prove Theorem \ref{thm: performance-gain}, we first analyze the expression of $|h_{r,1}+h_{r,2}e^{\jmath\frac{\pi}{2}}|^2 $, shown as follows
\begin{small}
    \begin{align}
    |h_{r,1}+h_{r,2}e^{\jmath\frac{\pi}{2}}|^2 
    &=\left|\sum_{n=1}^{N_{1}}|h_{1,n}||g_{1,n}|\right|^2+\left|\sum_{n=1}^{N_{2}}|h_{2,n}||g_{2,n}|\right|^2
   \nonumber \\ &
    \quad -\sqrt{2}\sum_{n=1}^{N_{1}}|h_{1,n}||g_{1,n}|\sum_{n=1}^{N_{2}}|h_{2,n}||g_{2,n}| \nonumber \\
    &\overset{a}{=}\! (N_{1}^{2}+N_{2}^{2}\!-\!\sqrt{2}N_{1}N_{2})\rho_{2}\rho_{3}  \!\overset{b}{=}\!\alpha_{1}\rho_{2}\rho_{3}N^{2} \nonumber
\end{align}
\end{small}
where ``$a$'' holds with assumption of LoS channel, ``$b$'' follows from the variable definition $\alpha_{1}=\zeta_{1}^{2}+\zeta_{2}^{2}-\sqrt{2}\zeta_{1}\zeta_{2}$ with $\zeta_{1}=\frac{\sqrt{3}+1-\sqrt{2}}{2}$, $\zeta_{2}= \frac{\sqrt{2}+1-\sqrt{3}}{2}$ from Theorem \ref{thm-elements-allocation}. Similarly, we have $|h_{r,2}|=\alpha_{2}\rho_{2}\rho_{3}N^{2}$ where $\alpha_{2}=\zeta_{2}^{2}$. Note that $h_{r,1}$ and $h_{r,2}$ can be obtained from Table \ref{table:Mapping-Euclidean-distance}.

With the tractable expressions derived above, the BERs of the primary and secondary signals under the proposed RIS design scheme can be respectively rewritten as
\begin{align}
    P_{s}^{\mathrm{pro}}&= \mathcal{Q}(\sqrt{2\alpha_{1}\rho_{2}\rho_{3}N^{2}}\mu)\overset{c}{=}\mathcal{Q}(\sqrt{\alpha_{1}\gamma_{b}N^{2}}), \\
    P_{c}^{\mathrm{pro}}&=\mathcal{Q}(\sqrt{2\alpha_{1}\rho_{2}\rho_{3}N^{2}}\mu)+\mathcal{Q}(2\sqrt{\alpha_{2}\rho_{2}\rho_{3}N^2}\mu)
   \nonumber \\ &
    \overset{d}{=} \mathcal{Q}(\sqrt{\alpha_{1}\gamma_{b}N^{2}})+ \mathcal{Q}(\sqrt{2\alpha_{2}\gamma_{b}N^{2}}),
\end{align}
where $\mu=\sqrt{\frac{p_{t}}{2\sigma^{2}}}$; ``$c$'' and ``$d$'' follow from $\gamma_{b}=\frac{p_{t}\rho_{2}\rho_{3}}{\sigma^{2}} $, which is defined as the SNR of the reflected link. With the fact that $\mathcal{Q}(\sqrt{\mathrm{SNR}})\leq \frac{1}{2}e^{-\mathrm{SNR}/2}$, the BERs upper bounds of $P_{s}^{\mathrm{pro}}$ and $P_{c}^{\mathrm{pro}}$ can be obtained. Then, by subtracting the BERs under the conventional RIS design from the obtained upper bounds, Theorem \ref{thm: performance-gain} immediately follows.
\end{appendices}
\vspace{-0.3cm}
\bibliographystyle{IEEEtran}
\bibliography{refFile}
\end{document}